\documentclass[a4paper,showpacs,superscriptaddress,10pt,nofootinbib]{revtex4-1}
\usepackage{amsmath,amssymb,bm,natbib}
\usepackage{epsfig}
\usepackage{graphicx,psfrag}
\usepackage{multirow}
\usepackage{slashed}
\usepackage{pstricks}
\usepackage{caption}
\usepackage{subcaption}
\makeatletter

\newcommand{\Rmnum}[1]{\expandafter\@slowromancap\romannumeral #1@}
\makeatother
\begin{document}
\newcommand{\ttbar}{$t\overline{t}$\,\,}
\newcommand{\ee}{$e^+e^-$}
\newcommand{\nc}{\newcommand}
\nc{\jc}{\frac{1}{4}}  \nc{\sll}{S_{LL}}     \nc{\slr}{S_{LR}}
\nc{\srl}{S_{RL}}      \nc{\srr}{S_{RR}}     \nc{\vll}{V_{LL}}
\nc{\vlr}{V_{LR}}      \nc{\vrl}{V_{RL}}     \nc{\vrr}{V_{RR}}
\nc{\tll}{T_{LL}}      \nc{\tlrs}{T_{LR}}    \nc{\trl}{T_{RL}}
\nc{\trr}{T_{RR}}      \nc{\slld}{S_{LL}^D}  \nc{\slrd}{S_{LR}^D}
\nc{\srld}{S_{RL}^D}   \nc{\srrd}{S_{RR}^D}  \nc{\vlld}{V_{LL}^D}
\nc{\vlrd}{V_{LR}^D}   \nc{\vrld}{V_{RL}^D}  \nc{\vrrd}{V_{RR}^D}
\nc{\tlld}{T_{LL}^D}   \nc{\tlrd}{T_{LR}^D}  \nc{\trld}{T_{RL}^D}
\nc{\trrd}{T_{RR}^D}   \nc{\aqde}{\alpha_{qde}}
\nc{\alq}{\alpha_{\ell q}}        \nc{\alqp}{\alpha_{\ell q'}}
\nc{\alqt}{\alpha_{\ell q}^{(3)}} \nc{\alqtc}{\alpha_{\ell
q}^{(3)*}} \nc{\alqj}{\alpha_{\ell q}^{(1)}}
\nc{\alqjc}{\alpha_{\ell q}^{(1)*}} \nc{\aeu}{\alpha_{eu}}
\nc{\alu}{\alpha_{\ell u}} \nc{\aqe}{\alpha_{qe}}
\nc{\ber}{\begin{eqnarray*}} \nc{\enr}{\end{eqnarray*}}
\nc{\jmpb}{(1-\beta)/(1+\beta)} \nc{\wspR}{r}      \nc{\varx}{x}
\nc{\bt}{\beta}

\nc{\non}{\nonumber} \nc{\lspace}{\;\;\;\;\;\;\;\;\;\;}
\nc{\llspace}{\lspace \lspace}
\nc{\jnl}{\frac{1}{{\mit\Lambda}^2}} \nc{\jd}{\frac{1}{2}}
\def\eett{e^+e^-\to t\bar{t}}

\def\tt{t\bar{t}}
\nc{\bb}{\bibitem} \nc{\ra}{\rightarrow} \nc{\g}{\gamma}
\nc{\beq}{\begin{equation}} \nc{\eeq}{\end{equation}}

\def\dps{\displaystyle}

\title{
Generalized top-spin analysis and new physics \\ in
$e^{+} e^{-}$ collisions with beam polarization}
\vspace*{1.5cm}
\author{{\bf B. Ananthanarayan }}
\author{{\bf Jayita Lahiri}} 
\author{{\bf Monalisa Patra}}
\affiliation{
Centre for High Energy Physics, 
Indian Institute of Science, 
Bangalore 560 012, India} 

\author{{\bf Saurabh D. Rindani}}

\affiliation{ 
Theoretical Physics Division,
Physical Research Laboratory,
Navrangpura, Ahmedabad 380 009, India}


\begin{abstract}
\vspace*{0.5cm}
A generalized top-spin analysis proposed some time ago in the context of Standard Model
and subsequently studied in varying contexts is now applied primarily to
the case of $e^+e^-\rightarrow t\bar{t}$ with transversely polarized beams.
This extends our recent work with new physics couplings of scalar ($S$) 
and tensor ($T$) types. We carry out a comprehensive 
analysis assuming only the electron beam to be transversely polarized, 
which is sufficient to
probe these interactions, and also eliminates any azimuthal angular dependence 
due to standard model or new physics of
vector ($V$) and axial-vector ($A$) type interactions. 
We then consider new physics of general four-Fermi 
type of $V$ and $A$ type with both
beams transversely polarized and discuss
implications with longitudinal polarization as well.
 The generalized spin bases are all investigated in the
presence of either longitudinal or transverse beam polarization to look for 
appreciable deviation from the
SM prediction in case of the new physics. 90\% confidence level limits are obtained 
on the interactions for the generalized spin bases with realistic integrated
luminosity. In order to achieve this we present a general discussion
based on helicity amplitudes and derive a general transformation matrix that enables us to treat the
spin basis. We find that beamline basis combined with transverse polarization provides
an excellent window of opportunity both for $S$, $T$ and $V$, $A$ new physics, followed by the off diagonal basis. 
The helicity basis is shown to be the best in case of longitudinal polarization to look for new physics effects
due to $V$ and $A$.

\end{abstract}

\pacs{14.65 Ha,13.66 Bc,13.88.+e}

\maketitle

\section{Introduction}

The International Linear Collider~\cite{ILCTDR} continues to be the foremost candidate
for testing Standard Model (SM) in the high energy frontier at high precision and is expected to
be the successor to the LHC in this regime.  The issue of beam
polarization has been an important subject for theorists and
experimentalists alike and pioneering studies have been carried out to
establish the importance of a physics programme based on the availability
of this~\cite{PolarizationReport}.  There are several choices that face designers, including that of
longitudinal or transverse beam polarization of one or both beams. In
particular, if transverse beam polarization of only one of the beams is
available, then any beyond the standard model (BSM) physics due to  vector
and axial-vector like interactions will not be visible at linear order in
the new physics.  It is only BSM physics due to scalar and tensor like
interactions that would be visible at linear order due to simple
considerations such as the chirality of interactions.

Our approach is based on the need to define a strong polarization programme through a set
of analytically accessible processes. 
At the planned ILC,  $e^{+} e^{-}\rightarrow t\bar{t}$ is 
a process that will be studied at great precision to validate the
SM and to look for deviations from it and is particularly suitable to meet this end. 
The process is of continued current interest, 
see e.g.,~\cite{hioki1},~\cite{hioki2},~\cite{hioki3},~\cite{hep-ph/0202045},~\cite{arXiv:1005.1756},~\cite{arXiv:1008.3562},
\cite{arXiv:1206.1033},~\cite{Willenbrock},~\cite{arXiv:1202.6659},~\cite{arXiv:1209.0547}
and references therein. At a linear collider the top quarks are produced in a unique spin configuration.
Since the top quark has a very short lifetime the definite spin state in which 
the $t\bar{t}$ pair is produced is not spoiled by hadronisation effects.
Due to this the direction of the spin of the top quark is reflected in the angular distribution of its decay
products. There are significant angular correlations between the decay products of the top quark
and its spin and also between the decay products of the top quark and top antiquark. New physics effects if
present in the production or decay mechanism will increase its sensitivity to spin correlation by modifying the
angular distribution of the decay products~\cite{hep-ph/9911249},~\cite{hep-ph/0002006},~\cite{hep-ph/0003294},
\cite{hep-ph/0009047},~\cite{hep-ph/0011173},\cite{arXiv:1202.2345}.

The availability of beam polarization will significantly
enhance the sensitivity to new physics (NP) provided the electron and
positron beams have transverse polarization (TP) or longitudinal
polarization (LP), each complementing the other, with
distinct prospects of obtaining very high
degree of polarization for both beams~\cite{PolarizationReport}.
In this work, we will consider such a scenario to find any azimuthal angle
($\phi$) dependence in $e^+e^-\to t\bar{t}$,  where $\phi$ is the 
azimuthal angle of one of the final particle.  The beam direction
is taken as the z-axis, and in the plane perpendicular to it,
the polarization direction of the electron (or positron) is
taken as the x-axis.  Note the approach here is complimentary
to that taken in~\cite{BAMPSDR}  where both beams were taken to be
polarized.  In order to really probe the extent to which the
new physics can be probed using an analytical approach, we now
extend our considerations to a general spin basis that was proposed
by Parke and Shadmi (PS)~\cite{Parke} in the context of unpolarized and longitudinally
polarized beams.  It was argued that two specific choices of
such a basis known as `beam-line' (BLB) and `off-diagonal' bases (ODB)
could be advantageous as far as increasing sensitivity is concerned. The work was further extended by~\cite{Kodaira} 
which shows that the above advantages for ODB will not change appreciably when QCD corrections are included. 
The impact of such bases in the
context of anomalous couplings of the top quarks and in case of various new physics models 
has been studied extensively in the literature~\cite{Nasuno},~\cite{Lin},~\cite{Wang}. The
main conclusion in those studies is that though the $t\bar{t}$ spin is most correlated in ODB 
compared to BLB and the helicity basis (HB) in SM, this enhanced spin correlation is not that
beneficial for distinguishing new physics effect in case of unpolarized or longitudinally polarized
beams.  To our knowledge the study of new physics along with the SM, has not been considered
in the context of transverse polarization. In order to meet this objective we present a discussion
on paving to these general spin bases in a formalism that employs known helicity amplitudes and a new method of introducing
a transformation.

  Whereas in reality, the top-quark spins are reconstructed only
from the decays, our treatment which is analytical cannot account
for decays since there is no clear cut closed form basis in which
the transverse beam polarization can be accounted for.  The inclusion of
decays being an important tool for spin analysis has been considered 
in the literature~\cite{correlation},~\cite{Sumino}.
However a direct analysis of the top quark spin structure itself is an insightful exercise.
The general spin basis considered here is a further diagnostic tool. By restricting ourselves to the
$t\bar{t}$ final state without the decays also allows us to describe analytically the initial transverse beam polarization,
since the kinematics continue to stay accessible. Thus we have a consistent framework of inclusion 
of initial beams TP effects and general spin basis of final state top quark pairs.  Nevertheless,
the objective of our work is to establish that both CP-violating
and CP-conserving BSM physics can be probed to linear order in
an effective manner with one beam being polarized and advantages
of a final-state spin analysis in a general basis.

We have employed in our study an effective field theory
approach to look for physics beyond the standard model in a model
independent manner. This is done by introducing higher dimensional operators
consistent with the symmetries of SM namely gauge invariance and Lorentz invariance~\cite{BW}.
Since these operators are of higher dimensions, by dimensional analysis their coefficients have inverse
powers of mass. The relevant higher dimensional operators, in the context of top pair
production are listed in~\cite{Grzad} and references therein. In this work
we have considered the non-standard interactions due to scalar and tensor type operators
which cannot be probed at linear order unless TP is available~\cite{BASDRPRD}, along with
the vector and axial-vector type operators.  These $V$, $A$ operators can be probed at linear
order with both unpolarized as well as longitudinal and transversely polarized beams. In this 
case both the beams have to have TP, to see their effect, unlike the scalar and tensor type
operators whose effect can be seen with one or both the beams having TP.
The question we pose is, how can the effect of these operators be tested and how do they
behave in different spin bases. The fact that spin reconstruction of the top quarks
is now feasible, and has been tested in hadron colliders~\cite{arXiv:1012.3093},~\cite{arXiv:1110.4194},~\cite{arXiv:1203.4081}
motivates this work to look for these NP effects in different bases in the case of TP.
Since we are looking for NP effects in the production mechanism, it is worth mentioning
that, it has been recently pointed out
that top polarization can be measured reliably from decay charged-lepton
angular distributions without errors arising from the $tbW$ 
couplings~\cite{hep-ph/0605100}.

The scheme of this paper is as follows:  In Sec.~\ref{np} we present our
formalism.  In Sec.~\ref{spin_basis} we recall for completeness the general spin basis.
In Sec.~\ref{rotation} we
present the distribution in case of transverse polarization in terms of helicity amplitudes
along with the transformation matrix that takes the helicity amplitudes from the helicity basis
to any generic spin basis. In Sec.~\ref{distribution} we
present the contribution of the BSM physics to the differential
distribution and discuss the C, P and T properties of the distribution. 
In Sec.~\ref{analysis} we carry out a numerical analysis with
some realistic choices of BSM parameters and demonstrate the
advantages of the general basis, in case of both TP and LP. We also
obtain 90\% confidence level (CL) limits on the couplings provided no
signal is observed for realistic beam polarization and integrated luminosity at the ILC.
In Sec.~\ref{conclusions} we present a discussion and our conclusions.

\section{Top-quark spin correlation at linear colliders}\label{np}

At future $e^+e^-$ linear colliders the spin of the top quark can be 
studied efficiently. This is due to the parity-violating interactions 
in top-quark production, which makes the produced tops
naturally polarized. Furthermore the polarization of the initial beams also 
helps in controlling the top polarization.

  In $e^+e^-$ collisions, $t\bar{t}$ is produced as follows:
  \begin{equation}
  e^+e^-\rightarrow V^* \rightarrow t\bar{t},~~V=\gamma,Z.
  \end{equation}
  
  The aim of this work is to look for new physics effect in the top production and its
effect in the spin correlation.  
  
\subsection{Four-Fermi operators}\label{four_fermi}  
  
The fact that SM 
describes physics well upto the electroweak symmetry breaking scale, it can be viewed as a low energy theory, with the heavy fields being integrated out.
Considering that new physics appears at a mass scale $\Lambda$, the Lagrangian can be written as an expansion in $1/\Lambda$:

\begin{equation}\label{lag}
{\cal L}_{eff}={\cal L}^{S\!M}+
\frac{1}{{\mit\Lambda}^2}\sum_i(\:\alpha_i{\cal O}_i+{\rm
h.c.}\:),
\end{equation}
where $\alpha$'s are the coefficients which parametrize
non-standard interactions(see ref.~\cite{BASDRPRD} and references therein).

The operators generated at the tree level, which will directly contribute to the production process
are as follows:

\begin{eqnarray}\label{operators}
&&{\cal O}_{lq}^{(1)}=\frac{1}{2}(\bar{l}\gamma_\mu l)(\bar{q}\gamma^\mu q),~~~ 
{\cal O}_{lq}^{(3)}=\frac{1}{2}(\bar{l}\gamma_\mu \tau^I l)(\bar{q}\gamma^\mu \tau^I q),~~~
{\cal O}_{eu} =\frac{1}{2}(\bar{e}\gamma_\mu e)(\bar{u}\gamma^\mu u), \nonumber \\
&&{\cal O}_{lu}=(\bar{l} u)(\bar{u} l),~~~
{\cal O}_{qe}=(\bar{q} e)(\bar{e} q),~~~ 
{\cal O}_{lq}=(\bar{l} e)\epsilon(\bar{q} u),~~~
{\cal O}_{lq'}=(\bar{l} u)\epsilon(\bar{q} e), 
\end{eqnarray}

where $l,q$ denote respectively the left-handed electroweak $SU(2)$
lepton and quark doublets, and $e$ and $u$ denote $SU(2)$ singlet
charged-lepton and up-quark right-handed fields. $\tau^I$ $(I=1,2,3)$ are
the Pauli matrices, and $\epsilon$ is the $2\times 2$ 
anti-symmetric matrix, $\epsilon_{12}=-\epsilon_{21}=1$, and
generation indices are suppressed.

The four Fermi operators listed above, containing the information about
physics beyond SM, after Fierz transformation takes the form
\begin{equation}\label{lag4f}
{\cal L}^{4F}
 =\sum_{i,j=L,R}\Bigl[\:S_{ij}(\bar{e}P_ie)(\bar{t}P_jt) + V'_{ij} (\bar{e}\gamma_\mu P_i e)(\bar{t}\gamma^{\mu}P_jt)
 +T_{ij}
 (\bar{e}\frac{\sigma_{\mu\nu}}{\sqrt{2}}P_ie)
(\bar{t}\frac{\sigma^{\mu\nu}}{\sqrt{2}}P_jt)\:\Bigr],
\end{equation}
with the coefficients satisfying the following constraints:
\begin{eqnarray}
S\equiv S_{RR}=S^{*}_{LL},\ \ \ S_{LR}=S_{RL}=0, \ \ \
T\equiv T_{RR}=T^{*}_{LL},\ \ \
T_{LR}=T_{RL}=0, \ \ \
V'\equiv V'_{ij}=V_{ij}^{'*}.
\end{eqnarray}
In Eq.~(\ref{lag4f}), $P_{L,R}$ are respectively the left- and right-chirality
projection matrices and as shown in~\cite{Grzad}, the $S_{ij}, V'_{ij}$ and
$T_{ij}$ can be expressed in terms of the $\alpha's$  from the four-Fermi part of the Lagrangian Eq.~(\ref{lag4f}). 
It can be checked that the Lagrangian in invariant under CP and T if 
$S$ and $T$ are real. In other words, non-zero values of Im~$S$ and/or Im~$T$ 
would imply CP violation. The $V',A'$ terms are invariant under CP and T. 

We have carried our analysis in terms of the helicity amplitudes given in the Appendix.~\ref{hel_amp}, which are same
as Ref.~\cite{Grzad}, with the normalization factor taken care of.

\section{Spin Bases}\label{spin_basis}
   
 The study of the polarized top quark requires a definite spin basis. It
has been shown by Parke and Shadmi~\cite{Parke},
that the degrees of spin correlations of top quarks in SM are quite
distinct in different bases.
The description of a generic spin basis by Parke and Shadmi
 \cite{Parke} is with  top
spin states defined in the top-quark rest frame,  with the
spin-quantization direction $\hat{s}_t$, making an angle  
$\xi^{PS}$ (we  use the label PS to distinguish it from the angle we
choose below) with the 
$\bar{t}$ measured in the direction of the incoming electron. The same definition holds for $\bar{t}$ spin states, 
with $\hat{s}_{\bar{t}}$, making an angle $\xi^{PS}$ with the $t$ momentum in 
the  direction of the incoming positron
in $\bar{t}$ rest frame. In our work $\xi$ is the angle measured, from the direction opposite
to the outgoing $\bar{t}$ in the direction of incoming electron, in $t$ rest frame.
Similar definition holds for the $\bar t$ rest frame, with $\xi$ measured 
from the direction opposite to the outgoing $t$ in the direction of incoming positron. 
For a polarized state $t_U\bar{t}_U$ refers to a top with spin along $\hat{s}_t$ and a 
top antiquark with spin along $\hat{s}_{\bar{t}}$. Analogous definitions
hold for the polarized states $t_D\bar{t}_D$, $t_U\bar{t}_D$ and
$t_D\bar{t}_U$

With the above convention for $\xi$, three different bases are constructed :
\begin{enumerate}
\item Helicity basis :
\begin{equation}\label{bas1}
\cos \xi = 1,
\end{equation}
with the top quark spin defined  along its direction of motion.
Moreover, in this basis, 
the polarized state $t_U\bar{t}_D$, will be the same as $t_R\bar{t}_L$
The partial analysis in this basis with both 
beams polarized and one of the top spins summed over was performed by us recently~\cite{BAMPSDR}. Here we are generalizing
that study. 
\item Beamline basis :
\begin{equation}\label{bas2}
\cos \xi=\frac{\cos \theta +\beta}{1+\beta \cos\theta},
\end{equation}
where $\beta$ is the top quark speed with $\beta = \sqrt{1-4m_t^2/s}$, and
$\theta$ is the top scattering angle with respect to the electron beam. In 
this basis the top-quark spin axis is the electron direction in the top rest frame
and the top anti-quark spin axis is the positron direction in the anti-top rest frame.
It may be checked that in the ultra relativistic limit ($\beta = 1$),
this basis reduces to the helicity basis.
\item Off-Diagonal basis :

For this basis it can be shown in SM, the
production cross sections of the like spin states $t_U\bar{t}_U$ and $t_D\bar{t}_D$
vanish for the left and right handed electron beam for particular values
of $\xi_L$ and $\xi_R$, given by
\begin{equation}\label{od1}
\cos \xi_L=\frac{(\beta  (V_{LL}-V_{LR})+\cos \theta
   (V_{LL}+V_{LR}))}{\sqrt{(\beta  (V_{LL}-V_{LR})+\cos
   \theta (V_{LL}+V_{LR}))^2+ \frac{4m_t^2}{s} \sin ^2 \theta
   (V_{LL}+V_{LR})^2}},
\end{equation}
\begin{equation}\label{od2}
\cos \xi_R=\frac{ (\beta  (V_{RR}-V_{RL})+\cos \theta
   (V_{RL}+V_{RR}))}{\sqrt{ (\beta  (V_{RR}-V_{RL})+\cos
   \theta (V_{RL}+V_{RR}))^2+\frac{4 m_t^2}{s} \sin ^2 \theta
   (V_{RL}+V_{RR})^2}},
\end{equation}
where
\begin{eqnarray}
&&V_{LL} = V^\gamma_{LL} + V^Z_{LL},~~~~~~V_{LR} = V^\gamma_{LR} + V^Z_{LR},\\ \nonumber
&&V_{RL} = V^\gamma_{RL} + V^Z_{RL},~~~~~V_{RR} = V^\gamma_{RR} + V^Z_{RR},
\end{eqnarray}
and
\begin{eqnarray}
&&V^{\cal V}_{LL}=(c^{e {\cal V}}_V + c^{e {\cal V}}_A) (c^{t {\cal V}}_V + c^{t {\cal V}}_A)/(s - M_{\cal V}^2),\\ \nonumber
&&V^{\cal V}_{LR} = (c^{e {\cal V}}_V + c^{e {\cal V}}_A) (c^{t {\cal V}}_V - c^{t {\cal V}}_A)/(s - M_{\cal V}^2), \\ \nonumber
&&V^{\cal V}_{RL} = (c^{e {\cal V}}_V - c^{e {\cal V}}_A) (c^{t {\cal V}}_V + c^{t {\cal V}}_A)/(s - M_{\cal V}^2), \\ \nonumber
&&V^{\cal V}_{RR} = (c^{e {\cal V}}_V - c^{e {\cal V}}_A) (c^{t {\cal V}}_V - c^{t {\cal V}}_A)/(s - M_{\cal V}^2), 
\end{eqnarray}
with ${\cal V} = \gamma, Z$ and
\begin{eqnarray}
&&c^{e Z}_V = (-1/2 + 2 s^2_w)/(2 \sqrt{1 - s^2_w} \sqrt{s^2_w}),~~~c^{e Z}_A = -1/(4 \sqrt{1 - s^2_w} \sqrt{s^2_w}), \\ \nonumber
&&c^{t Z}_V = (1/2 - 4 s^2_w/3)/(2 \sqrt{1 - s^2_w} \sqrt{s^2_w}),~~~c^{t Z}_A = 1/(4 \sqrt{1 - s^2_w} \sqrt{s^2_w}),  
\end{eqnarray}

\end{enumerate}
 where $s^2_w = \sin^2 \theta_W$ = 0.231, the Weinberg angle and 
$m_t$ = 172 GeV,  $m_Z$ = 90.1 GeV is the mass of the top and $Z$ boson respectively.

There are two off-diagonal bases for the pair production considered here, one for
$e^-_Le^+_R$ and the other for $e^-_Re^+_L$ scattering. For the $t\bar{t}$ production 
the two bases are almost coincident since the ratio $V_{LL}/V_{LR}$ in Eq.~(\ref{od1}) is approximately equal in sign and
magnitude to the ratio  $V_{RR}/V_{RL}$ in Eq.~(\ref{od2}). Therefore for the rest of our numerical analyses we have used
$\cos \xi_L$, henceforth written as $\cos \xi$ as the ODB. It is worth mentioning that ODB approaches the helicity
basis for $\beta \rightarrow$ 1.
                  We would further like to mention that in our work we will call the ODB as the
standard model off diagonal basis (SMOD). Before embarking on to BSM physics we need 
to consider whether or not in the SM in the presence of TP this basis defined by 
Eqs.~(\ref{od1}),~(\ref{od2}) continue to have the desirable property of $t_U\bar{t}_U$ and $t_D\bar{t}_D$
vanishing or not. An inspection of the defining condition given in Eq. (11) of PS and the 
distribution in the presence of TP shows that the property of SMOD where the final state with like 
spin configuration vanishes holds. 
Thus in the presence of TP, the choice for $\cos\xi$
given by Eqs.~(\ref{od1}),~(\ref{od2}) continues to be SMOD. Thus any contribution to the cross section of $t_U\bar{t}_U$ 
and $t_D\bar{t}_D$ with and without TP is a signal of BSM physics.

Since for $\beta \rightarrow$ 1, the BLB and SMOD approach the HB our definition of the 
convention for $\xi$ gives a self-consistent set of bases.

\section{Transverse Polarization and General Spin Basis}\label{rotation}

   The different spin bases described in the previous section have been 
investigated in case of SM~\cite{Parke}. It is found for the off-diagonal basis
in polarized $e^+e^-$ colliders not only do the like spin configurations vanish,
but one spin configuration dominates the total cross-section. The behaviour
of the bases for SM in the case of TP has not been addressed before.
There has been a study to explore which spin basis is more suitable for studying new physics
effects in top quark production ~\cite{Lin},~\cite{Wang} in the presence of LP. Here we consider the
new physics effects described in SubSection~\ref{four_fermi}, containing scalar and tensor type interactions along with
non-standard vector and axial-vector type of interactions. In earlier works ~\cite{BASDRPRD},~\cite{BAMPSDR} it was
shown that scalar and tensor type operators
cannot be probed at linear order unless TP is available.  The cross section in the presence of TP, with new physics
effects is calculated from the helicity amplitudes from the expression~\cite{Hikasa},
where contributions proportional to $|T_{LLIJ}|^2$, $|T_{RRIJ}|^2 $ and $T_{RRIJ}^* T^{}_{LLIJ}$
are discarded.

\begin{eqnarray}\label{tp_dist}
\frac{d\sigma (e^+e^-\rightarrow t_I\bar{t}_J)}{d\cos\theta d\phi}&=&
|T_{RLIJ}|^2+|T_{LRIJ}|^2-\frac{1}{2}P^T_{e^-}P^T_{e^+}Re~e^{-2i\phi}T^*_{RLIJ}T^{}_{LRIJ} \nonumber \\
&& +\frac{1}{2}P^T_{e^-} Re~e^{-i\phi} (T^*_{RLIJ}T^{}_{LLIJ}+T^*_{RRIJ}T^{}_{LRIJ}) \nonumber \\
&& -\frac{1}{2}P^T_{e^+}Re~e^{-i\phi}(T^*_{RLIJ}T^{}_{RRIJ}+T^*_{LLIJ}T^{}_{LRIJ}).
\end{eqnarray}

In the above equation, $P^T_{e^-}$ and $P^T_{e^+}$ are the degree of TP of the
electron and positron respectively. The direction of the electron polarization
is fixed along the positive $x$ axis, with the azimuthal angle of the polarization vector 
to be zero in the c.m. frame. Moreover we have also considered the polarization vectors
of the electron and the positron in the opposite direction.

The helicity amplitudes used for our analysis are defined in the Appendix.~\ref{hel_amp},
with the same order.
In the expressions above, ${IJ}$ denotes the different final-state spin combinations of $UD$, $DU$, $UU$ and $DD$.  
The above expression is in the HB,
therefore in order to study the TP effect in other bases a rotation is performed on the spin of the top and antitop 
as described in the next subsection.

\subsection{The transformation matrix}
The amplitudes in a generic spin basis may be obtained from the
amplitudes in the helicity basis by means of a transformation
corresponding to a rotation of the $t$ and $\bar t$ spin bases. 
We are giving here the expressions for a more general case, by considering different
rotation angles $\xi_t$ and $\xi_{\bar t}$ in the spin space of the $t$ and $\bar t$. Thus we
can write

\beq
\left( 
\begin{array}{c}
T'_{LRUU} \\T'_{LRUD}\\  T'_{LRDU}\\ T'_{LRDD}\\
 T'_{RLUU}\\ T'_{RLUD}\\ T'_{RLDU}\\ T'_{RLDD} \\ \end{array} \right) = R'(\xi_t, \xi_{\bar t}) \left( 
\begin{array}{c}  T^{}_{LRUU} \\ T^{}_{LRUD}\\  T^{}_{LRDU}\\ T^{}_{LRDD}\\
 T^{}_{RLUU}\\ T^{}_{RLUD}\\ T^{}_{RLDU}\\ T^{}_{RLDD}     \\ 
\end{array}
\right),
\eeq
where the left-hand side corresponds to the helicity conserving amplitudes in a generic
spin basis, and $LR \rightarrow LL$, $RL \rightarrow RR$ for the helicity violating amplitudes.
The amplitudes $T$ on the right-hand side are in the
helicity basis, and $R'(\xi_t, \xi_{\bar t})$ is the transformation matrix corresponding
to the parameters $\xi_t$ and $\xi_{\bar t}$. $R'(\xi_t, \xi_{\bar t})$ operates on a column vector spanned by
$U$ and $D$ in the helicity basis and is related to the $4\times4$ matrix  $M(\xi_t, \xi_{\bar t})$ by,

\beq 
R(\xi_t, \xi_{\bar t}) = \left( \begin{array}{cc} \label{equ:rot}
M(\xi_t, \xi_{\bar t}) & 0 \\ 0 & M(\xi_t, \xi_{\bar t}) 
\end{array} \right),
\eeq

where $M(\xi_t, \xi_{\bar t})$ is the direct product of two rotation
matrices. $R_t(\xi_t)$ and $R_{\bar t}(\xi_{\bar t})$ parametrize the effect of rotation in the spin space of $t$ and $\bar t$,
respectively. In a schematic notation, we have

\beq
R_t(\xi_t)= \left( 
\begin{array}{cc} 
\cos\frac{\xi_t}{2} & -\sin\frac{\xi_t}{2} \\
\sin\frac{\xi_t}{2} & \cos\frac{\xi_t}{2} 
\end{array}
\right),
\eeq

and analogously

\beq
R_{\bar t} (\xi_{\bar t}) = \left( 
\begin{array}{cc} 
\cos\frac{\xi_{\bar t}}{2} & -\sin\frac{\xi_{\bar t}}{2} \\
\sin\frac{\xi_{\bar t}}{2} & \cos\frac{\xi_{\bar t}}{2} 
\end{array}
\right).
\eeq

Thus, we have
$M \equiv R_t(\xi_t) \otimes R_{\bar t} (\xi_{\bar t})$
given by 
\newcommand{\xibytwo}{\frac{\xi}{2}}
\beq
M(\xi_t, \xi_{\bar t}) = \left( \begin{array}{cccc}\label{mat1}

\cos\frac{\xi_t}{2} \cos\frac{\xi_{\bar t}}{2} & -\cos\frac{\xi_t}{2} \sin\frac{\xi_{\bar t}}{2}
 &  -\sin\frac{\xi_t}{2}\cos\frac{\xi_{\bar t}}{2}  &\sin\frac{\xi_t}{2} \sin\frac{\xi_{\bar t}}{2}  \\
\cos\frac{\xi_t}{2}\sin\frac{\xi_{\bar t}}{2} & \cos\frac{\xi_t}{2}\cos\frac{\xi_{\bar t}}{2}  
& -\sin\frac{\xi_t}{2} \sin\frac{\xi_{\bar t}}{2} & -\sin\frac{\xi_t}{2}\cos\frac{\xi_{\bar t}}{2}\\
\sin\frac{\xi_t}{2} \cos\frac{\xi_{\bar t}}{2}& - \sin\frac{\xi_t}{2}\sin\frac{\xi_{\bar t}}{2}
& \cos\frac{\xi_t}{2} \cos\frac{\xi_{\bar t}}{2}  & - \cos\frac{\xi_t}{2}\sin\frac{\xi_{\bar t}}{2}\\
\sin\frac{\xi_t}{2}\sin\frac{\xi_{\bar t}}{2}  & \sin\frac{\xi_t}{2} \cos\frac{\xi_{\bar t}}{2} 
& \cos\frac{\xi_t}{2}\sin\frac{\xi_{\bar t}}{2} & \cos\frac{\xi_t}{2}\cos\frac{\xi_{\bar t}}{2} 

\end{array}
\right) .
\eeq

However in the present work we take $\xi_t = \xi_{\bar t} = \xi$.
Note however that we have the possibility of an even further generalization
when $\xi_t \neq \xi_{\bar t}$, which is not studied here. The matrix defined in Eq.~(\ref{mat1}) has the property that when one goes to
the generalized spin basis, e.g., the $(LRDU)$ amplitude
in that basis gets an admixture from the $(LRUU)$, $(LRUD)$ and $(LRDD)$ of
the helicity basis. Similar argument applies for the other helicity amplitudes
in the generic spin basis. After taking $\cos\xi$ =1 for the helicity basis,
the $IJ$ indices in the original and rotated bases are (trivially) the
same. The matrix reduces to identity matrix for $\cos \xi = 1$. The value of $\cos \xi$ for the different
bases are defined in Eqs.~(\ref{bas1}), (\ref{bas2}),~(\ref{od1}) and (\ref{od2}).
In case of SMOD with the choice of angle $\cos \xi_L$ the helicity
amplitudes  $T'_{LRUU}$ and $T'_{LRDD}$ vanish, whereas for the choice $\cos \xi_R$, the helicity 
amplitudes  $T'_{RLUU}$ and $T'_{RLDD}$ vanish. 
Our choice of the angle $\xi$ described in Sec.~\ref{spin_basis} thus makes
transparent why the HB and the ODB are so called as exemplified in the general derivation.
 We note that upon $\phi$ integration of Eq.~(\ref{tp_dist})
we obtain the results of  PS~\cite{Parke} except of a factor of 2. Our result is larger by
a factor of 2 and when we sum over all the helicities of the final state our result agrees with the SM
prediction~\cite{BASDRPRD}.

\section{Distributions and their properties} \label{distribution}    
      
      We present here the distribution in the presence of TP, for
different new physics of the type $S$, $T$, $V$ and $A$.  The degree of polarization 
expected in ILC is about 80\% for the electron and 60\% for the positrons~\cite{PolarizationReport}. 
In the distributions given below, 
$e^+e^-\rightarrow t_U\bar{t}_U~ or~ t_D\bar{t}_D$ is defined as $UU/DD$ and $e^+e^-\rightarrow t_U\bar{t}_D~ or~ t_D\bar{t}_U$ is defined as $UD/DU$
and
\begin{eqnarray}
A_L&=&V_{LL}+V_{LR},~~A_R=V_{RL}+V_{RR}\nonumber \\
B_L&=&V_{LL}-V_{LR},~~B_R=V_{RL}-V_{RR}\nonumber \\
B&=&\frac{\alpha}{2}\sqrt{\frac{3\beta}{s}},~~ A=\sqrt{\frac{3\beta}{64\pi^2s}} \label{eq:albl}.
\end{eqnarray}

The differential cross section for the process $e^-e^+ \rightarrow t\bar{t}$, in presence
of the new physics $S$ and $T$ in generic spin bases
for different spin configurations is obtained by:

\begin{eqnarray}\label{eqn_stuu}
\frac{d\sigma_{UU/DD}}{d\cos\theta\ d\phi}&=&\frac{d\sigma_{SM}}{d\cos\theta\ d\phi}
+\frac{1}{8} A B s \left( \sin \phi \left\lbrace\sqrt{s} \sin \xi [(A_L-A_R) \cos \theta +\beta  (B_L+B_R)] \right. \right. \nonumber \\
&&\left. \left. +2 m_t (A_R-A_L) \cos \xi \sin \theta \right\rbrace \left\lbrace(P^T_{e^-}-P^T_{e^+}) 
\left[\sqrt{s} \cos \xi (2 \rm{Im}T \cos \theta -\beta  \rm{Im}S) \right. \right. \right. \nonumber \\
&& \left. \left. \left. + 4 m_t \rm{Im}T \sin \xi \sin \theta \right]\mp \sqrt{s} 
(P^T_{e^-}+P^T_{e^+})(\rm{Im}S-2 \beta  \rm{Im}T \cos \theta)\right\rbrace \right.  \nonumber \\
&& \left. +\cos \phi \left\lbrace \sqrt{s} \sin \xi [(A_L+A_R) \cos \theta +\beta  
(B_L-B_R)]-2 m_t (A_L+A_R)\cos \xi \sin \theta \right\rbrace \right.  \nonumber \\
&& \left. \times \left\lbrace  (P^T_{e^-}+P^T_{e^+}) \left[\sqrt{s} \cos \xi 
(2 \rm{Re}T \cos \theta -\beta \rm{Re}S)+4 m_t \rm{Re}T \sin \xi \sin \theta \right]\right. \right.  \nonumber \\
&& \left.\left. \pm \sqrt{s} (P^T_{e^+}-P^T_{e^-}) (\rm{Re}S-2 \beta  \rm{Re}T\cos \theta)\right\rbrace 
+  2(P^T_{e^-}+P^T_{e^+}) \left\lbrace \sqrt{s} \sin \xi (A_L \cos \theta +\beta  B_L)\right.\right. \nonumber \\
&&\left. \left. -2 A_L m_t \cos \xi \sin \theta \right\rbrace 
\left\lbrace \sqrt{s} \left[\cos \xi \cos \phi ~(\beta  \rm{Re}S-2 \rm{Re}T \cos \theta)\pm \sin \phi
(\rm{Im}S-2 \beta \rm{Im}T \cos \theta)\right]\right.\right. \nonumber \\
&&\left.\left. -4 m_t \rm{Re}T \sin \theta \sin \xi \cos \phi \right\rbrace  \right),
\end{eqnarray}

where      

\begin{eqnarray}\label{sm_uu}
\frac{d\sigma_{SM}}{d\cos\theta\ d\phi}&=& \frac{1}{16} B^2 s \left(s \sin ^2 \xi \left[\left(A_L^2+A_R^2\right) 
\cos ^2\theta +2 \beta  \cos
\theta (A_L B_L-A_R B_R)+\beta ^2 \left(B_L^2+B_R^2\right)\right] \right. \nonumber \\
&& \left. -4 m_t \cos \xi \sin \theta\left(\sin \xi \sqrt{s} \left[\left(A_L^2+A_R^2\right) \cos \theta +\beta 
(A_L B_L-A_R B_R)\right] - m_t \cos \xi \sin \theta \left(A_L^2+A_R^2\right)\right)\right. \nonumber \\
&&\left. +2 s P^T_{e^-} P^T_{e^+} \cos 2 \phi \left[2 A_L \frac{m_t}{\sqrt{s}} \cos \xi \sin \theta - \sin \xi (\beta  B_L+A_L \cos \theta)\right]\right. \nonumber \\
&&\left. \times \left[2 A_R \frac{m_t}{\sqrt{s}} \cos \xi \sin \theta+\sin \xi (\beta  B_R-A_R \cos \theta) \right]\right),
\end{eqnarray}

and

\begin{eqnarray}\label{eqn_stud}
\frac{d\sigma_{UD/DU}}{d\cos\theta\ d\phi}&=&\frac{d\sigma^{'}_{SM}}{d\cos\theta\ d\phi}
+\frac{1}{8} A B s \left((P^T_{e^-}-P^T_{e^+}) \sin \phi \left\lbrace 4 m_t \rm{Im}T \cos \xi \sin \theta \right. \right. \nonumber \\
&& \left. \left.   +\sqrt{s} \sin \xi(\beta  \rm{Im}S-2 \rm{Im}T \cos \theta)\right\rbrace \left\lbrace \sqrt{s}
\cos \xi [(A_L-A_R) \cos \theta +\beta (B_L+B_R)]\right. \right. \nonumber \\
&& \left. \left.  \pm[A_L+A_R+\beta  (B_L-B_R)\cos \theta] + 2 m_t (A_L-A_R) 
\sin \xi \sin \theta \right\rbrace \right. \nonumber \\
&&\left.   +(P^T_{e^-}+P^T_{e^+}) \cos \phi \left\lbrace 4 m_t \rm{Re}T \cos \xi 
\sin \theta +\sqrt{s} \sin \xi (\beta  \rm{Re}S-2 \rm{Re}T \cos \theta)\right\rbrace \right. \nonumber \\
&&\left.  \times \left\lbrace \sqrt{s} \cos \xi [(A_L+A_R) \cos \theta +\beta  (B_L-B_R)]
\pm[A_L-A_R+\beta (B_L+B_R) \cos \theta] \right. \right. \nonumber \\
&&\left. \left.  +2 m_t (A_L+A_R) \sin \xi \sin \theta \right\rbrace -2(P^T_{e^-} +P^T_{e^+}) \cos \phi
\left\lbrace \sqrt{s} \left[\cos \xi (A_L \cos \theta +\beta  B_L)\right.\right.\right. \nonumber \\
&&\left. \left. \left. \pm A_L\pm \beta  B_L \cos \theta \right]+2 A_L m_t \sin \theta  \sin \xi \right\rbrace
\left\lbrace \sqrt{s} \sin \xi (\beta \rm{Re}S-2 \rm{Re}T \cos \theta) \right. \right.  \nonumber \\
&&\left. \left. +4 m_t \rm{Re}T \cos \xi \sin \theta \right\rbrace \right),
\end{eqnarray}
      
where      

\begin{eqnarray}\label{sm_ud}
\frac{d\sigma^{'}_{SM}}{d\cos\theta\ d\phi}&=&\frac{1}{32} B^2 s \left(s \cos \xi \left\lbrace 2\cos \xi \left[\left(A_L^2+A_R^2\right)  \cos ^2 \theta
+2 \beta  \cos \theta (A_L B_L-A_R B_R)+ \beta ^2
\left(B_L^2+B_R^2\right)\right]\right. \right. \nonumber \\
&&\left. \left.   +4 \cos \theta \left[\pm(A_L^2-A_R^2) \pm \beta ^2 (B_L^2-B_R^2)\right]\pm4 \beta  (\cos ^2 \theta +1) (A_L B_L+A_R B_R)\right\rbrace \right. \nonumber \\
&&\left.   +2s \left[ \left(A_L^2+A_R^2\right)+2 \beta  \cos \theta (A_L B_L-A_R B_R)+ \beta ^2 \cos ^2 \theta
\left(B_L^2+B_R^2\right) \right]\right. \nonumber \\
&&\left.   +8 m_t \sqrt{s} \sin \xi \sin \theta  \left\lbrace  \cos \xi\left[\left(A_L^2+A_R^2\right)
 \cos \theta +\beta (A_L B_L-A_R B_R)\right] \right.\right. \nonumber \\
&&\left.\left. \pm (A_L^2-A_R^2) +\beta  \cos\theta (A_L B_L+A_R B_R)\right\rbrace  
+ 8 m_t^2 \left(A_L^2+A_R^2\right) \sin ^2 \xi \sin ^2 \theta  \right. \nonumber \\
&&\left.  +4s P^T_{e^-} P^T_{e^+} \cos 2 \phi \left\lbrace \pm \cos \xi (A_L \cos \theta 
+\beta  B_L)+A_L+\beta  B_L\cos \theta \pm 2 A_L  \frac{m_t}{\sqrt{s}} \sin \xi \sin \theta  \right\rbrace  \right.  \nonumber \\
&&\left. \times \left\lbrace -(A_R \pm \beta  B_R \cos \xi)
+ \cos \theta (\pm A_R \cos \xi +\beta  B_R) \pm 2 A_R \frac{m_t} {\sqrt{s}} \sin \xi \sin \theta  \right\rbrace \right).
\end{eqnarray}

The distribution in presence of the new physics of the vector and axial-vector type  
denoted as $A^{'}_L$, $A^{'}_R$, $B^{'}_L$, $B^{'}_R$, with the same form as $A_L$, $A_R$, $B_L$, $B_R$ defined
in Eq.~(\ref{eq:albl}) is obtained by:

\begin{eqnarray}\label{eqn_vauu}
\frac{d\sigma_{UU/DD}}{d\cos\theta\ d\phi}&=&\frac{d\sigma_{SM}}{d\cos\theta\ d\phi} 
+ \frac{d\sigma_{NP}}{d\cos\theta\ d\phi}  + \frac{d\sigma_{NP}^{TP}}{d\cos\theta\ d\phi}.
\end{eqnarray}

The contributions to
$A^{'}_{L,R}$ and $B^{'}_{L,R}$ enter as vertex corrections to the $t\bar{t}\gamma$ and $t\bar{t}Z$ vertices and from some
additional gauge bosons due to some higher symmetry suppressed by the new physics scale.
In the  equation above $\frac{d\sigma_{SM}}{d\cos\theta\ d\phi}$ is the SM distribution
given in Eq.~(\ref{sm_uu}), $\frac{d\sigma_{NP}}{d\cos\theta\ d\phi}$ is the distribution
in the absence of TP in case of new physics given by:

\begin{eqnarray}
\frac{d\sigma_{NP}}{d\cos\theta\ d\phi}&=&
\frac{1}{8} B^2 s \left(-2 m_t \sqrt{s} \cos \xi \sin \xi \sin \theta  
\left\lbrace \beta  (A_L B^{'}_L+A^{'}_L B_L-A_R B^{'}_R-A^{'}_R B_R) \right. \right. \nonumber \\
&& \left. \left.   +2 \cos \theta (A_L A^{'}_L+A_R A^{'}_R)\right\rbrace + s\sin ^2 \xi 
\left\lbrace \beta  \cos \theta (A_L B^{'}_L+A^{'}_L B_L-A_R B^{'}_R-A^{'}_R B_R)\right. \right. \nonumber \\
&& \left. \left.    +\cos ^2 \theta (A_L A^{'}_L+A_R A^{'}_R)+\beta ^2 (B_L B^{'}_L+B_R B^{'}_R)\right\rbrace 
+4 m_t^2 \cos ^2 \xi \sin ^2 \theta  (A_L A^{'}_L+A_R A^{'}_R)\right),
\end{eqnarray}

and $\frac{d\sigma_{NP}^{TP}}{d\cos\theta\ d\phi}$ shows the distribution due to new physics in the presence
of TP:

\begin{eqnarray}
\frac{d\sigma_{NP}^{TP}}{d\cos\theta\ d\phi}&=&  \frac{1}{8}B^2 s P^T_{e^-} P^T_{e^+}  
\cos 2 \phi  \left( -2m_t \sqrt{s} \cos \xi \sin \xi \sin \theta 
\left[\beta  (-A_L B^{'}_R-A^{'}_L B_R+A_R B^{'}_L+A^{'}_R B_L)\right.\right.\nonumber \\
&&\left. \left.  +2 \cos \theta (A_L A^{'}_R+A^{'}_L A_R)\right]+s \sin ^2 \xi 
\left[ \beta \cos \theta (-A_L B^{'}_R-A^{'}_L B_R+A_R B^{'}_L+A^{'}_R B_L)\right.\right.\nonumber \\
&& \left. \left.   + \cos ^2 \theta (A_L A^{'}_R+A^{'}_L A_R) - \beta ^2 
(B_L B^{'}_R+B^{'}_L B_R)\right]
  +4 m_t^2 \cos ^2 \xi \sin ^2 \theta (A_L A^{'}_R+A^{'}_L A_R)\right).
\end{eqnarray}

Similarly
\begin{eqnarray}\label{eqn_vaud}
\frac{d\sigma_{UD/DU}}{d\cos\theta\ d\phi}&=&\frac{d\sigma^{'}_{SM}}{d\cos\theta\ d\phi}
+ \frac{d\sigma^{'}_{NP}}{d\cos\theta\ d\phi}  + \frac{d\sigma_{NP}^{TP'}}{d\cos\theta\ d\phi},
\end{eqnarray}
where the same definition follows as in Eq.~(\ref{eqn_vauu}) with  $\frac{d\sigma^{'}_{SM}}{d\cos\theta\ d\phi}$ given by Eq.~(\ref{sm_uu}).

\begin{eqnarray}
\frac{d\sigma^{'}_{NP}}{d\cos\theta\ d\phi}&=&\frac{1}{16} B^2 s \left(\cos \xi \left\lbrace \pm 4 m_t \sqrt{s} \sin \xi \sin \theta  
\left[\beta  (A_L B^{'}_L+A^{'}_L B_L-A_R B^{'}_R-A^{'}_R B_R) \right. \right. \right. \nonumber \\
&&\left. \left. \left.   +2 \cos \theta (A_L A^{'}_L+A_R A^{'}_R)\right]\pm 2s\left[2 \cos \theta 
\left(A_L A^{'}_L-A_RA^{'}_R+\beta ^2 (B_L B^{'}_L-B_RB^{'}_R)\right)\right. \right. \right. \nonumber \\
&& \left. \left. \left.  \pm \beta  (\cos ^2 \theta +1) (A_LB^{'}_L+A^{'}_L B_L+A_R B^{'}_R+A^{'}_RB_R)\right]\right\rbrace \right. \nonumber \\
&& \left.   +2s \cos ^2\xi \left\lbrace  \beta  \cos \theta (A_L B^{'}_L+A^{'}_L B_L-A_RB^{'}_R-A^{'}_R B_R)\right.\right. \nonumber \\
&& \left.\left. +\cos ^2 \theta (A_LA^{'}_L+A_R A^{'}_R) + \beta ^2 (B_L B^{'}_L+B_R B^{'}_R)\right\rbrace \right. \nonumber \\
&& \left. \pm 4m_t \sqrt{s} \sin \xi \sin \theta  \left\lbrace \beta  \cos \theta
(A_L B^{'}_L+A^{'}_L B_L+A_RB^{'}_R+A^{'}_R B_R)+2 (A_L A^{'}_L- A_RA^{'}_R)\right\rbrace \right. \nonumber \\
&& \left.   +2s \left\lbrace  \beta  \cos \theta (A_LB^{'}_L+A^{'}_L B_L-A_R B^{'}_R-A^{'}_RB_R)+ (A_L A^{'}_L+ A_R A^{'}_R) \right. \right.\nonumber \\
&& \left.  \left. + \beta ^2 \cos ^2\theta (B_L B^{'}_L+B_R B^{'}_R)\right\rbrace +8 m_t^2 \sin ^2 \xi \sin ^2 \theta (A_L A^{'}_L+A_R A^{'}_R)\right),
\end{eqnarray}
and
\begin{eqnarray}
\frac{d\sigma_{NP}^{TP'}}{d\cos\theta\ d\phi}&=&\frac{1}{8} B^2 sP^T_{e^-} P^T_{e^+}  
\cos 2 \phi \left( 2m_t \sqrt{s} \cos \xi \sin \xi \sin \theta 
\left[\beta  (-A_L B^{'}_R-A^{'}_L B_R+A_RB^{'}_L+A^{'}_R B_L)\right.\right.\nonumber \\
&& \left. \left.   +2 \cos \theta  (A_LA^{'}_R+A^{'}_L A_R)\right]+s \cos ^2 \xi \left[\beta 
\cos \theta (-A_L B^{'}_R-A^{'}_L B_R+A_RB^{'}_L+A^{'}_R B_L)\right.\right.\nonumber \\
&& \left.  \left.+\cos ^2 \theta (A_LA^{'}_R+A^{'}_L A_R)- \beta ^2 (B_L B^{'}_R+B^{'}_L B_R)\right]\right.\nonumber \\
&& \left. \mp \beta \sin \theta (s \cos \xi \sin \theta \mp m_t \sqrt{s} \sin \xi \cos \theta)
 (A_LB^{'}_R+A^{'}_L B_R+A_R B^{'}_L+A^{'}_RB_L)\right.\nonumber \\
&& \left.  +4 m_t^2 \sin^2 \xi \sin ^2 \theta (A_L A^{'}_R+A^{'}_L A_R) +s
\left[ \beta  \cos \theta (A_L B^{'}_R+A^{'}_LB_R-A_R B^{'}_L-A^{'}_R B_L)\right.\right.\nonumber \\
&& \left.  \left.- (A_LA^{'}_R+A^{'}_L A_R)+\beta ^2 \cos ^2 \theta (B_LB^{'}_R+B^{'}_L B_R)\right]\right).
\end{eqnarray}

It is interesting to examine the above distributions from the point of view of the C, P and 
T properties of the interactions. As noted earlier, the only couplings which
can lead to CP violation are Im~$S$ and Im~$T$. If, as noted 
in ref. \cite{BASDRPRD},
the $t$ and $\bar t$ spins are not observed, the only CP-odd observable 
possible is $(\vec p_{e^-}-\vec p_{e^+})\times (\vec s_{e^-} - \vec s_{e^+})
\cdot (\vec p_{t}- \vec p_{\bar t})$, which would get an expectation value 
from the terms in the distribution proportional to $(P^T_{e^-} + P^T_{e^-})\,
\sin\theta \cos\phi$. These terms can indeed be seen to be proportional to 
Im~$S$ or Im~$T$ in the sum of the distributions for the various $t$ and 
$\bar t$ up and down spin projections. When the $t$ and $\bar t$ spins are
observed, there are more observables possible which are CP odd, which depend on 
these spins. These are $(\vec p_{e^-}-\vec p_{e^+})\times (\vec s_{e^-} + 
\vec s_{e^+}) \cdot (\vec s_{t}- \vec s_{\bar t})$  and $(\vec p_{e^-}-
\vec p_{e^+})\times (\vec s_{e^-} - \vec s_{e^+})\cdot (\vec s_{t} + 
\vec s_{\bar t})$. 

We first take the simplest case of helicity basis 
distributions. Because of the dependence on the difference of $t$ and $\bar t$ 
spins in the first of these observables, it would occur in the difference 
between the 
distributions for the $UD$ and $DU$ spin projections. It can be seen to be 
proportional to $P^T_{e^-} - P^T_{e^+}$. The second observable
would occur in the sum of the distributions for the spin projections $UU$ and
$DD$, and would be proportional to $P^T_{e^-} + P^T_{e^+}$. In either case,
the angular dependence of the CP-violating terms will be proportional to 
$\sin\phi$, occurring with the couplings Im~$S$ and Im~$T$.

Let us now consider other spin bases. The amplitudes corresponding to each of these 
is obtained by transformation of the spin amplitudes by the matrix 
given of Eq.~(\ref{mat1}), with an appropriate value of $\xi$. Such a 
transformation, however, does not change the fact that the distributions,
to linear order in the new-physics couplings, contain the CP-violating couplings
Im~$S$ and Im~$T$ with the same azimuthal dependence, viz., $\sin\phi$. Thus,
in any spin basis, an asymmetry (which we discuss below) 
that can isolate the $\sin\phi$ terms, will be
a measure of CP violation. Also, the transformation to a different spin
basis does not change the dependence on $e^+$ and $e^-$ polarizations.

\section{Numerical Analysis}\label{analysis}

 We consider the azimuthal distribution of the final state, in the presence of different types
of non-standard couplings. In case of the $S$ and $T$ type interactions apart from the azimuthal distributions
different asymmetries are constructed to isolate their contributions. The results are presented for
the three different bases considered here. The effect of LP is also considered 
in the presence of the $V$ and $A$ type interactions,
on the total production cross section along with the fraction of the top quarks produced. An asymmetry is
also considered to test these interactions for different cases of initial LP, to measure the angular correlation of $t\bar t$.
We have also done an analysis to put bound on the various anomalous couplings considered here.

\subsection{New Physics due to $S$ and $T$ interactions}

\begin{figure}[htb]
\begin{subfigure}{.45\linewidth}
\centering
\includegraphics[width=7.5cm, height=6cm]{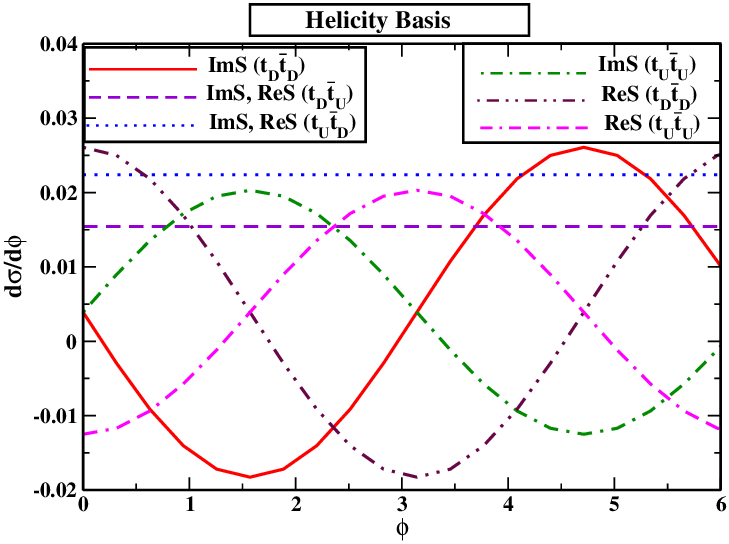}
\caption{}
\label{fig:hel_s}
\end{subfigure}%
\begin{subfigure}{.45\linewidth}
\centering
\includegraphics[width=7.5cm, height=6cm]{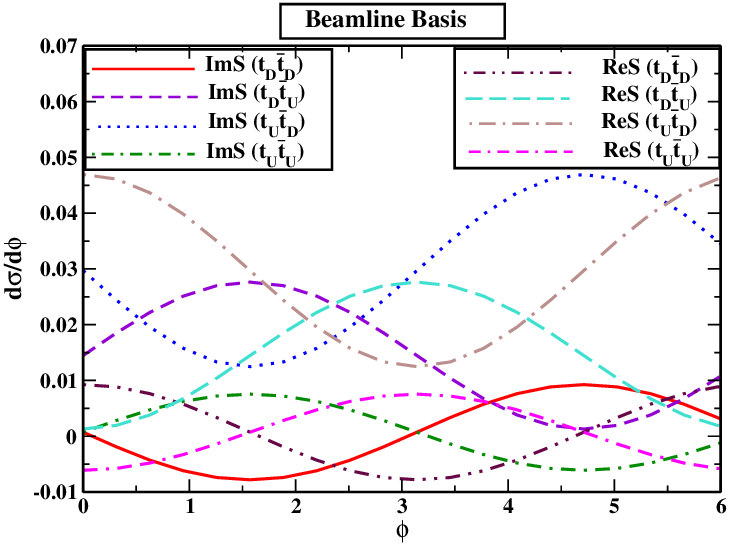}
\caption{}
\label{fig:beam_s}
\end{subfigure}\\[1ex]
\begin{subfigure}{\linewidth}
\centering
\vspace*{0.45cm}
\includegraphics[width=7.5cm, height=6cm]{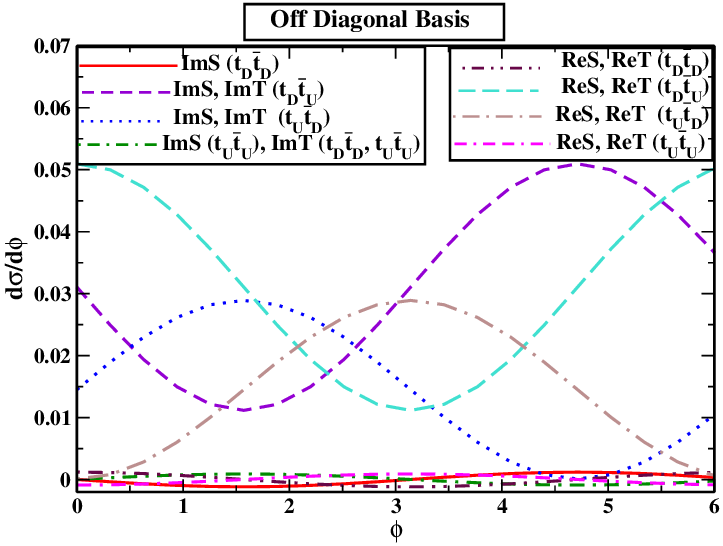}
\caption{}
\label{fig:off_s}
\end{subfigure}
\caption{The azimuthal distribution of the top quark pairs at a 500 GeV linear collider in case of different
bases considered here, with $P^T_{e^-}$ = 0.8 and $P^T_{e^+}$ = 0. The distribution in the presence of different anomalous couplings are considered,
with the value of the respective coupling to be 0.5 $\times$ 10$^{-6}$ GeV$^{-2}$, keeping the value of others zero.}
\label{fig:diff_bases}
\end{figure}

        Firstly the scalar and the tensorial type of couplings are considered.  We are considering the case with only one of the
initial beams being transversely polarized.  For our analysis we have taken $P^T_{e^-}$ = 0.8 and $P^T_{e^+}$ = 0. 
This choice of beam polarization has the advantage of eliminating the $\phi$ contribution from SM,
along with the contribution if any from the new physics due to $V$ and $A$ type of interactions.  The $\phi$ distribution in the
HB, BLB and SMOD is shown in Fig.~\ref{fig:diff_bases}. For the purpose of illustration we have taken magnitude of the NP
couplings to be of the order 10$^{-6}$ GeV$^{-2}$, inspired by the sensitivity that was expected at ILC with realistic polarization and
integrated luminosity~\cite{BAMPSDR}. In this work, the values of the couplings are chosen to be 0.5 $\times$ 10$^{-6}$ GeV$^{-2}$,
a choice for which we have checked that the linear approximation holds good. 

\vspace{0.2cm}

We note that the Figures given in this Subsection are plotted for two cases, ($a$) \rm{Re}$S$ = 0.5 $\times$ 10$^{-6}$ GeV$^{-2}$,
keeping the values of other couplings to be zero, and ($b$) \rm{Im}$S$ = 0.5 $\times$ 10$^{-6}$ GeV$^{-2}$,
with the other couplings equal to zero.
 The Fig.~\ref{fig:hel_s}, shows that in the HB for the final-state spin configuration of the form
$t_D\bar{t}_U$ and $t_U\bar{t}_D$,  \rm{Im}$S$ and \rm{Re}$S$ has no $\phi$ dependence. There is no
contribution to the $\phi$ dependence in this basis as $S$ does not contribute to the helicity amplitudes in
the $t_D\bar{t}_U$ and $t_U\bar{t}_D$ sector. However in the other bases Figs.~\ref{fig:beam_s},~\ref{fig:off_s}
due to the action of the transformation matrix Eq.~(\ref{equ:rot}), there are non trivial contributions.
The azimuthal distributions due to the presence of $T$ is almost the same as the
distribution from $S$, so we do not show their distribution in the HB and BLB. The contribution
due to $T$ is shown for the off diagonal basis, Fig.~\ref{fig:off_s}, with the contribution from either
\rm{Re/Im}$T$ to be  0.5 $\times$ 10$^{-6}$ GeV$^{-2}$, keeping the other couplings to be zero.
In case of SMOD, the distribution from the same final-state spin configuration $t_U\bar{t}_U$ and $t_D\bar{t}_D$ is almost equal to zero,
in the presence of $S$ and $T$. This behaviour is similar to the distribution in the presence of SM only. But 
the other final-state spin configurations show a measurable $\phi$ distribution contrary to SM behaviour which is
$\phi$ independent. For completion, we show in Table~\ref{tab_sm}, the contribution from SM in different bases, with different final-state
spin configuration. We would  like to point out that, provided in ILC only one of the beams is transversely polarized
observation of modulation in any of the final-state spin configuration will be a signature of $S$ and $T$ interactions.
This behaviour holds for all the three bases discussed here.
\begin{center}
\begin{table}
\begin{tabular}{||c|c|c|c||} \hline
Spin Configurations &Helicity Basis &Beamline Basis &SM Off Diagonal Basis \\ \hline \hline
~~~~~~~~~$t_D\bar{t}_D$ &0.0078 &0.0015 &0.0000 \\
~~~~~~~~~$t_D\bar{t}_U$ &0.0309 &0.0289 &0.0621 \\
~~~~~~~~~$t_U\bar{t}_D$ &0.0448 &0.0594 &0.0290 \\
~~~~~~~~~$t_U\bar{t}_U$ &0.0078 &0.0015 &0.0000 \\ \hline
\end{tabular}
\caption{The $\phi$ independent contribution coming from SM, in different bases with only one of the initial beams
being transversely polarized. This constant term is present in all the distribution due to $S$ and $T$ interactions.}
\label{tab_sm}
\end{table}
\end{center}
  The azimuthal distribution considered above has a supplement in the form of a constant term from SM.
We therefore consider various asymmetries which isolate the
contributions from $S$ and $T$ type of interactions only. 
The asymmetries considered here are those that were earlier studied
in~\cite{BAMPSDR}. The asymmetries are schematically given by :
\begin{eqnarray}
&& A_1(\theta)=\frac{1}{\sigma^{SM}(\theta)}\biggl[\int^\pi_0\frac{d\sigma}{d\Omega}
\,d\phi-\int^{2\pi}_\pi\frac{d\sigma}{d\Omega}\,d\phi\biggr]\label{asym1} \\
&& A_2(\theta)=\frac{1}{\sigma^{SM}(\theta)}\biggl[\int^{\frac{\pi}{2}}
_{-\frac{\pi}{2}}\frac{d\sigma}{d\Omega}\,d\phi-\int^{\frac{3\pi}{2}}_{\frac{\pi}{2}}\frac{d\sigma}{d\Omega}\,d\phi\biggr] \label{asym2}
\end{eqnarray}

where d$\sigma$ corresponds to a particular final state spin configurations as given in the Figures.

\vspace{0.2cm}
\begin{figure}[htb]
\begin{subfigure}{.45\linewidth}
\centering
\includegraphics[width=7.5cm, height=6cm]{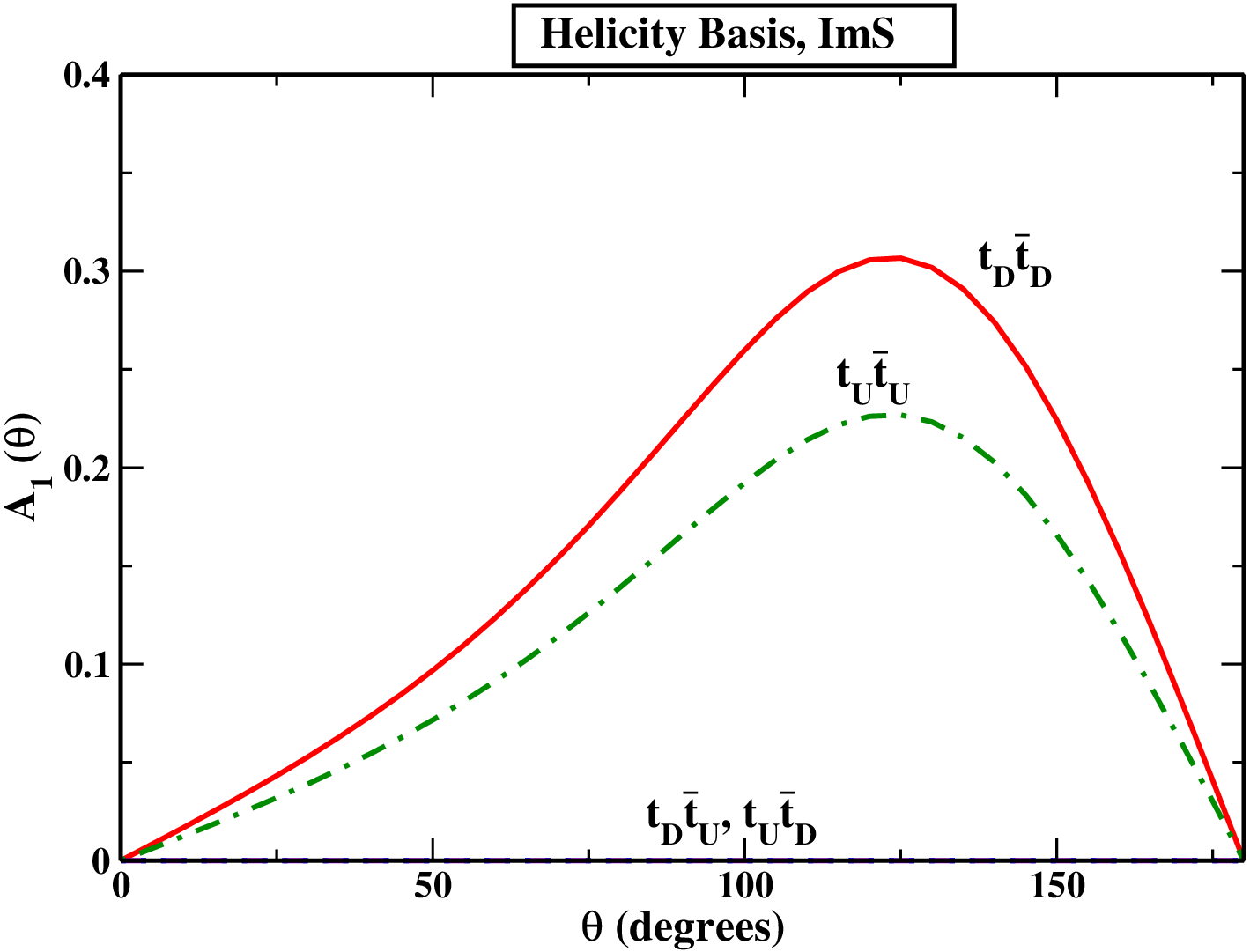}
\caption{}
\label{fig:asy1_hels}
\end{subfigure}%
\begin{subfigure}{.45\linewidth}
\centering
\includegraphics[width=7.5cm, height=6cm]{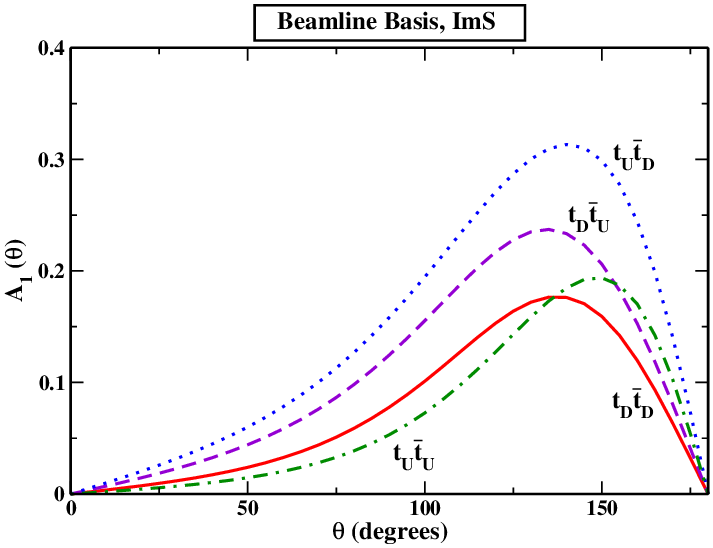}
\caption{}
\label{fig:asy1_beams}
\end{subfigure}\\[1ex]
\begin{subfigure}{\linewidth}
\centering
\vspace*{0.45cm}
\includegraphics[width=7.5cm, height=6cm]{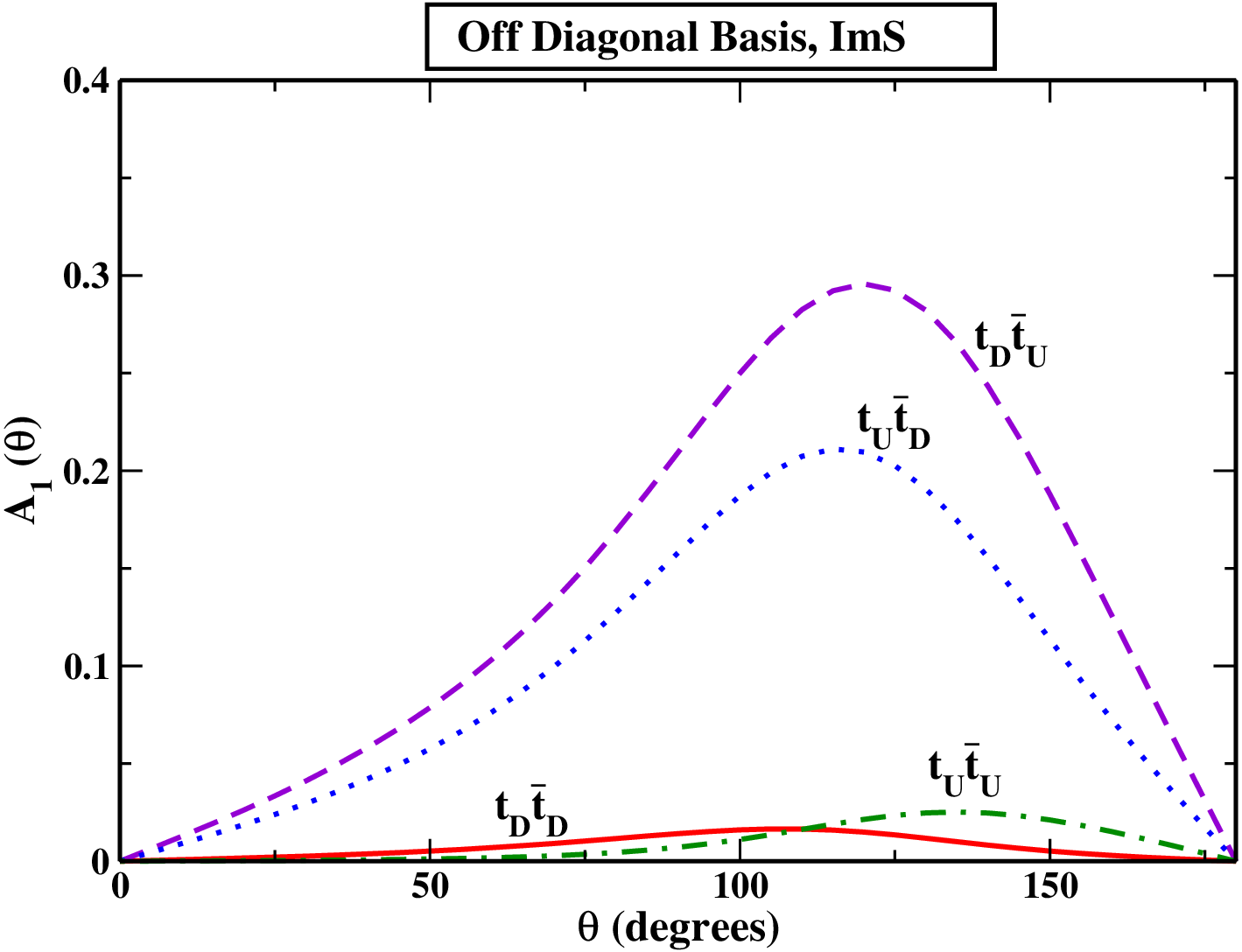}
\caption{}
\label{fig:asy1_offs}
\end{subfigure}
\caption{$A_1 (\theta)$ as a function of $\theta$ with $P^T_{e^-}$ = 0.8, $P^T_{e^+}$ = 0 at 
$\sqrt{s}$ = 500 GeV in case of different bases for all combinations of final-state spin 
configurations with $\rm{Im}S = 0.5 \times 10^{-6}$ GeV$^{-2}$, keeping the values of other couplings to be zero.}
\label{fig:asy1_all}
\end{figure}

The asymmetry defined in Eq.~(\ref{asym1}) contains both \rm{Im}$S$ and \rm{Im}$T$ with zero 
contribution from the real part of the couplings, in case of all the final-state spin configurations.  
In view of the earlier discussion, this asymmetry isolates the
CP-violating couplings. It may be noted, however, that the initial state
is not an eigenstate of CP, which would require the $e^+$ and $e^-$
polarizations to be equal in magnitude. The asymmetry, thus, is not
explicitly a
purely CP-odd observable. However, since it depends only on the
CP-violating couplings \rm{Im}$S$ and \rm{Im}$T$, it is a measure of CP
violation. 
Similarly the other asymmetry defined in Eq.~(\ref{asym2}) receives contribution from \rm{Re}$S$ and \rm{Re}$T$ only.  
It is thus a measure of CP-conserving interactions.

\begin{figure}[htb]
\begin{subfigure}{.45\linewidth}
\centering
\includegraphics[width=7.5cm, height=6cm]{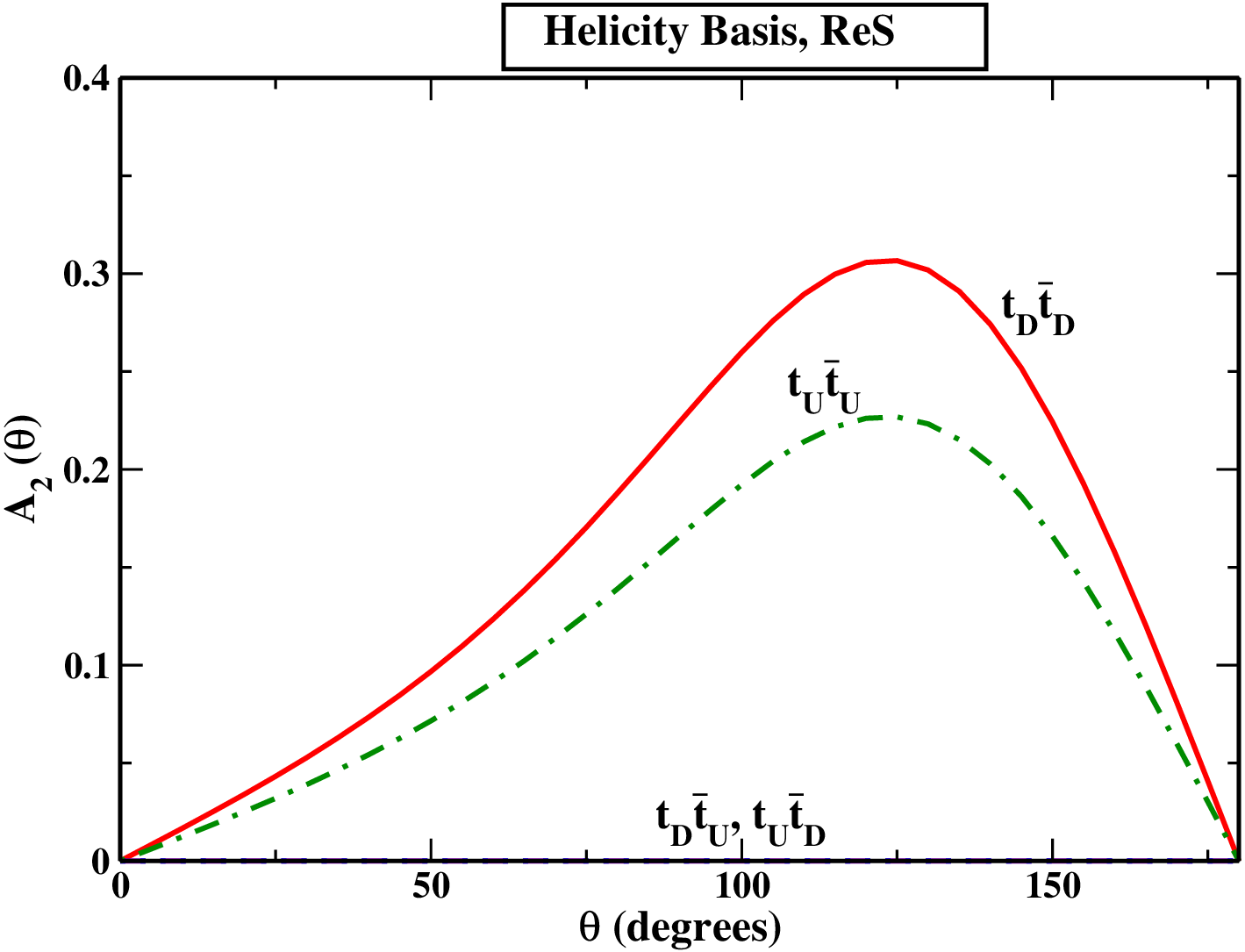}
\caption{}
\label{fig:asy2_hels}
\end{subfigure}%
\begin{subfigure}{.45\linewidth}
\centering
\includegraphics[width=7.5cm, height=6cm]{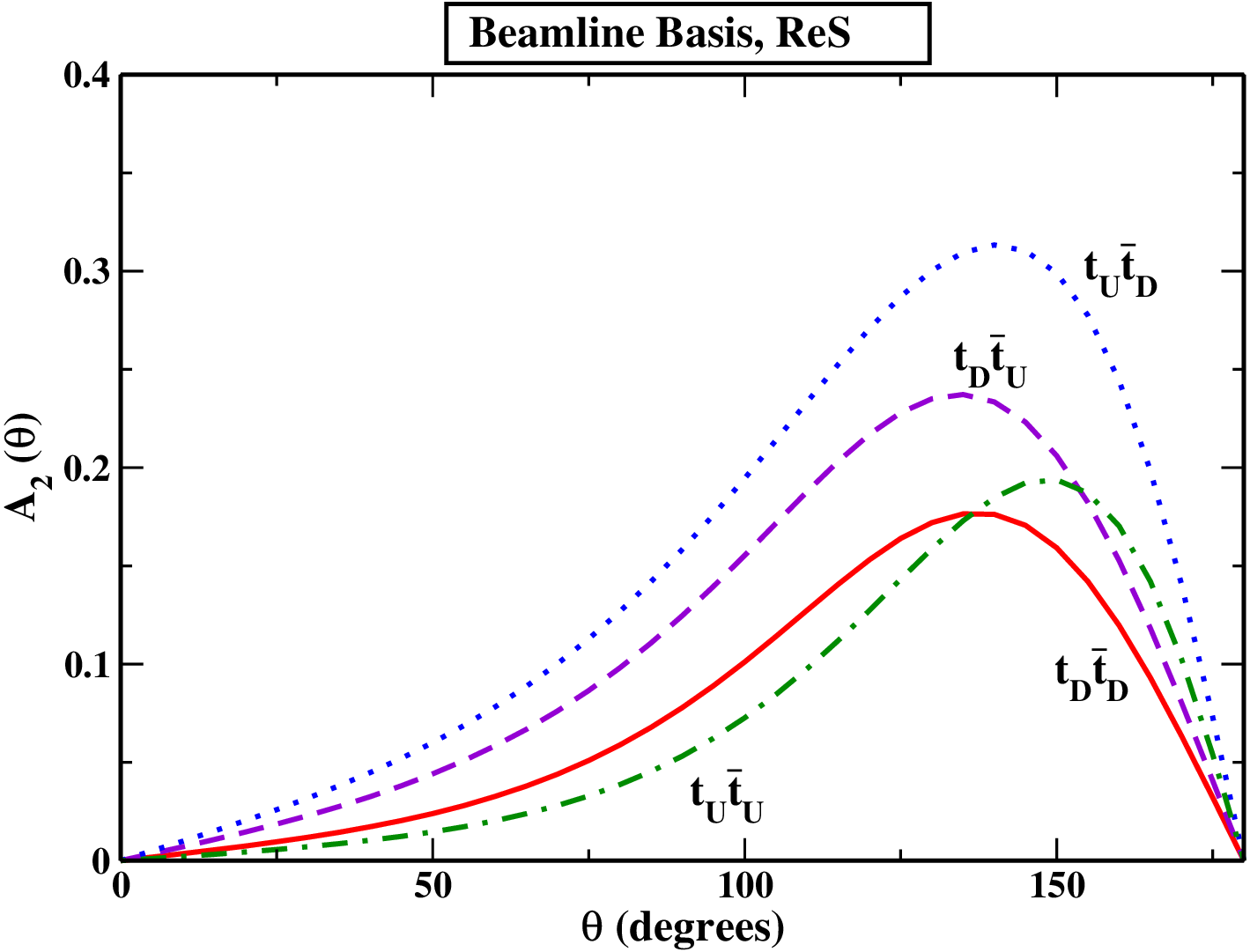}
\caption{}
\label{fig:asy2_beams}
\end{subfigure}\\[1ex]
\begin{subfigure}{\linewidth}
\centering
\vspace*{0.45cm}
\includegraphics[width=7.5cm, height=6cm]{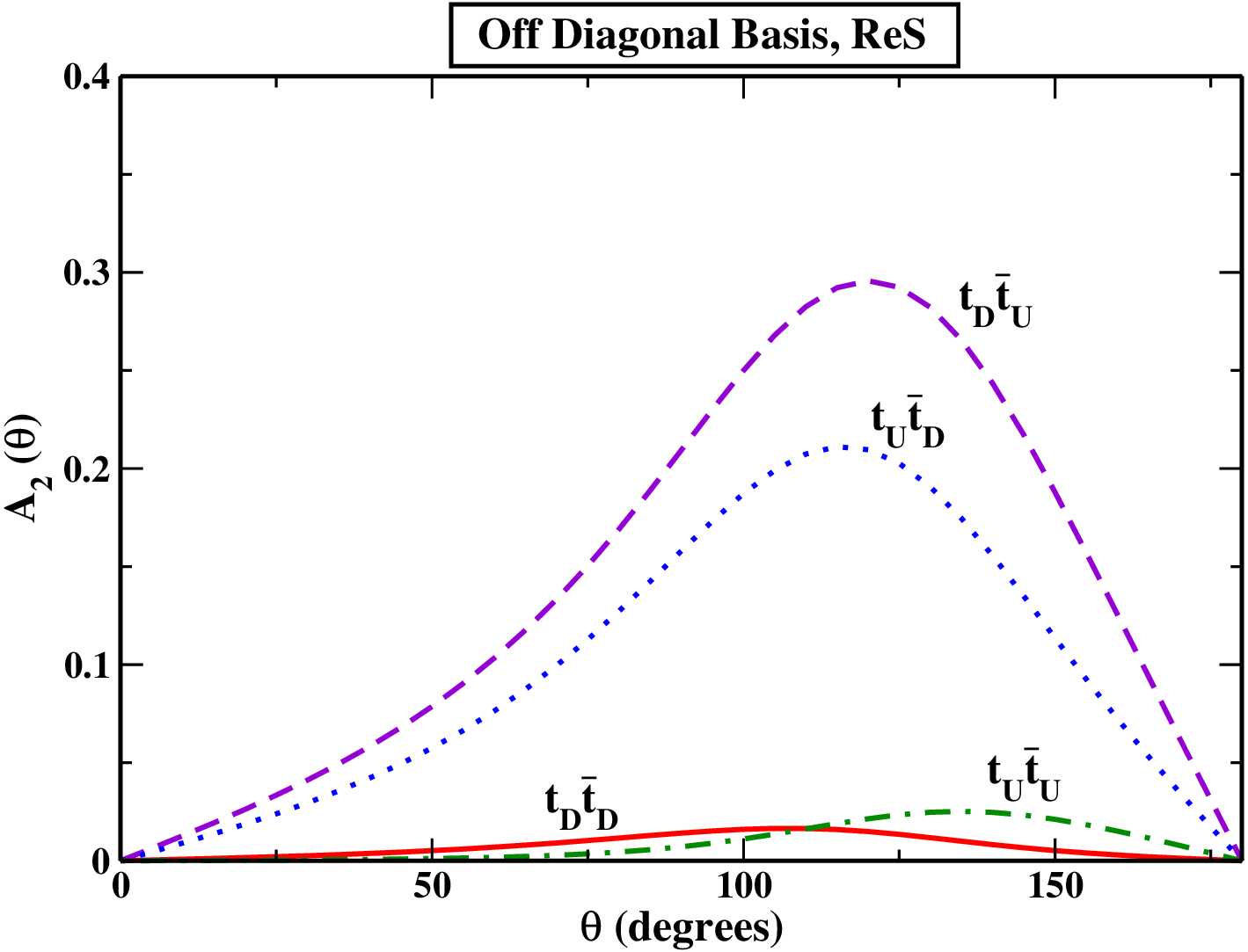}
\caption{}
\label{fig:asy2_offs}
\end{subfigure}
\caption{$A_2 (\theta)$ as a function of $\theta$ with $P^T_{e^-}$ = 0.8, $P^T_{e^+}$ = 0 at 
$\sqrt{s}$ = 500 GeV in case of different bases for all combinations of 
final-state spin configurations with $\rm{Re}S = 0.5 \times 10^{-6}$ GeV$^{-2}$ and the values of other couplings are
considered zero.}
\label{fig:asy2_all}
\end{figure}

For our calculations we have only considered the contribution from \rm{Im}$S$ and \rm{Re}$S$, with the value of
$T$ taken to be zero. In Figs.~\ref{fig:asy1_all} and \ref{fig:asy2_all} we show the asymmetry $A_1(\theta)$
and $A_2(\theta)$ as a function of $\theta$ at $\sqrt{s}$ = 500 GeV for all the bases. 
Depending on the bases involved, the new physics effect due to $S$ and $T$ contributes
to the asymmetries. Let us now discuss what can be concluded from Figs.~\ref{fig:asy1_all} and \ref{fig:asy2_all}.
From Fig.~\ref{fig:asy1_all} it is seen that for the various final-state spin combinations the HB, BLB and SMOD perform
almost similarly as regards the sensitivity to \rm{Im}$S$. The asymmetry for all the spin
configurations is most significant in BLB. The same result holds for \rm{Im}$T$. Fig.~\ref{fig:asy2_all},
also shows that \rm{Re}$S$ produces a similar signal as Im $S$.
We note that the observance of these asymmetries, in case of any of the beams being
transversely polarized will be a signal of $S$ and $T$ type of physics. We further note that in the presence of TP
the BLB is almost equally sensitive to effects from NP, for the different spin configurations of $t$ and $\bar t$
compared to SMOD and HB, which are sensitive to only  particular spin configurations. The largest asymmetries
are all comparable in the three bases.

\vspace{0.2cm}

It is also possible to ask what are the 90\% CL limits on $S$ and $T$ that can be obtained at the ILC.
For this purpose we consider an integrated luminosity 
of $\mathcal{L}$ = 500 fb$^{-1}$ at $\sqrt{s}$ = 500 GeV and the same beam polarization.
The sensitivity, characterized by the limit $\mathcal{C}_{limit}$ on a given
coupling, is given by 

\begin{equation}\label{limit}
\mathcal{C}_{limit} =\frac{1.64}{|A|\sqrt{N_{SM}}},
\end{equation}

where $|A|$ is the asymmetry for unit value of the coupling and $N_{SM}$ is the number of SM events. 
The asymmetries used are those given in Eqs.~(\ref{asym1}),(\ref{asym2}). However we consider the case of resolving the
$t$ spin and sum over the $\bar t$ spin and take the difference as given below 

\begin{equation}
\frac{d\sigma}{d\Omega}={\left.\frac{d\sigma}{d\Omega}\right|}_{UD+UU} 
-{\left. \frac{d\sigma}{d\Omega}\right|}_{DU+DD}
\end{equation}

\begin{figure}[htb]
\begin{subfigure}{.45\linewidth}
\centering
\includegraphics[width=7.5cm, height=6cm]{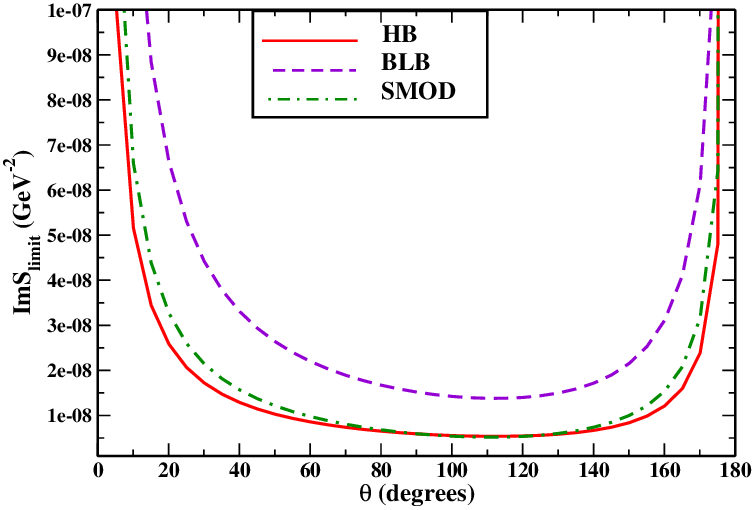}
\caption{}
\label{fig:ims_lim}
\end{subfigure}%
\begin{subfigure}{.45\linewidth}
\centering
\includegraphics[width=7.5cm, height=6cm]{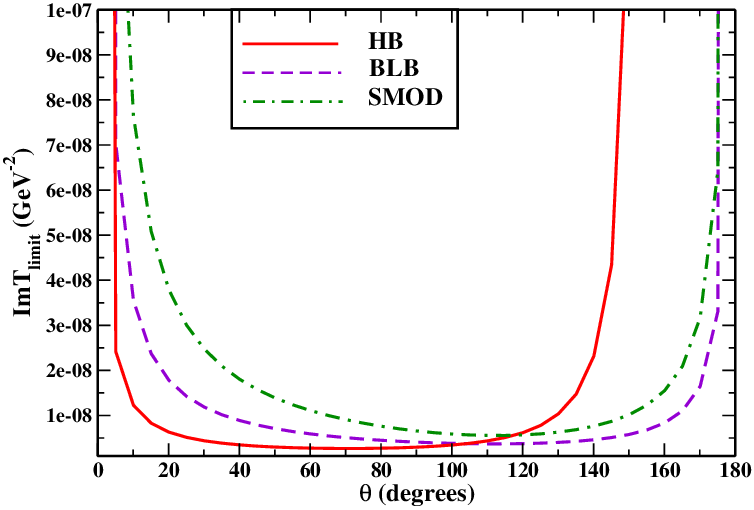}
\caption{}
\label{fig:imt_lim}
\end{subfigure}
\caption{90 \% C.L. limit obtained on the couplings Im$S$ and Im$T$ from
the modified asymmetry $A_1(\theta)$ 
at $\sqrt{s}$ = 500 GeV with an integrated luminosity of 500 fb$^{-1}$ for $P^T_{e^-}$ = 0.8 and $P^T_{e^+}$ = 0, plotted as a function of $\theta$ 
for different bases.}
\label{fig:limit}
\end{figure}

Similar analyses can be done by considering the spin of $\bar{t}$ and summing over the spin of $t$.
This is analogous to the considerations of~\cite{BAMPSDR} for each of the spin bases. However in~\cite{BAMPSDR}
both beams were perfectly polarized and only the HB was considered.
This is a continuation of that work to check the sensitivity of the different bases for obtaining limits
on $S$ and $T$. 
The asymmetries now are sensitive to the $t$ polarization dependent part of the cross section
and the number of events increases compared to the case when individual spin is measured.   We present the limits on the couplings Im$S$ and Im$T$,
for different bases in Fig.~\ref{fig:limit} for $P^T_{e^-}$ = 0.8 and $P^T_{e^+}$ = 0. The limits for 
Re$S$ and Re$T$ obtained from the modified asymmetry $A_2 (\theta)$ are similar to those obtained 
for the imaginary parts of those couplings so we do 
not present the result here. Fig.~\ref{fig:ims_lim},
shows that the best limit for Im$S$ is obtained from the HB and SMOD. The best limit is around $5\times10^{-9}$ GeV$^{-2}$
and is obtained at $\theta$ = 110$^\circ$.
The limit obtained from BLB is poorer by about an order of magnitude. Note that the limits obtained here depends on
the degree of TP. In the ideal condition $P^T_{e^-}, P^T_{e^+}$ = 1, the best limit is obtained
from the HB and BLB for Im$S$ and is around $3\times10^{-9}$ GeV$^{-2}$. SMOD fares badly in this case, and the limit obtained is
of the order $10^{-8}$ GeV$^{-2}$. Similarly for Im$T$, from Fig.~\ref{fig:imt_lim} we see that the best limit is obtained
from the HB and is $3\times10^{-9}$ GeV$^{-2}$. The other bases behave similarly and give a limit of $5\times10^{-9}$ GeV$^{-2}$. 
Here also the limits obtained from different bases are sensitive to the degree of TP. Therefore an investigation in different bases
with different degrees of TP is necessary to obtain limits on $S$ and $T$ type couplings.

\vspace*{0.65cm}

\subsection{New Physics due to $V$ and $A$ interactions}

   The analytical form of the differential distribution in the presence of transverse polarization, due to vector and axial-vector type 
of non-standard interactions is shown in Eqs.~(\ref{eqn_vauu}) and~(\ref{eqn_vaud}). These anomalous couplings are sensitive 
to longitudinal beam polarization, 
unlike the $S$ and $T$ interactions considered before. Moreover 
 their effect in the presence of transversely polarized beams can be seen only when both the beams are polarized.
We will first study their effect in the presence of TP, and compare the deviation from the SM. The case of unpolarized and longitudinal
beam polarization is considered later. 
\vspace{0.4cm}
\begin{figure}[htb]
\begin{subfigure}{.45\linewidth}
\centering
\psfrag{SM}[b][b][0.75]{\bf{SM}}
\psfrag{SM + AL}[b][b][0.75]{\bf{SM +} $\bf{A^{'}_{L}}$}
\psfrag{SM + AR}[b][b][0.75]{\bf{SM +} $\bf{A^{'}_{R}}$}
\includegraphics[width=7.5cm, height=6cm]{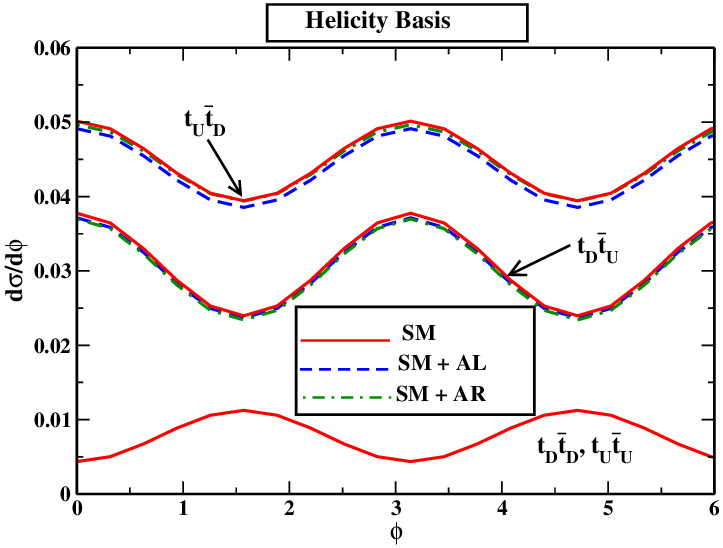}
\caption{}
\label{fig:hel_al}
\end{subfigure}%
\begin{subfigure}{.45\linewidth}
\centering
\psfrag{SM}[b][b][0.75]{\bf{SM}}
\psfrag{SM + AL}[b][b][0.75]{\bf{SM +} $\bf{A^{'}_{L}}$}
\psfrag{SM + AR}[b][b][0.75]{\bf{SM +} $\bf{A^{'}_{R}}$}
\includegraphics[width=7.5cm, height=6cm]{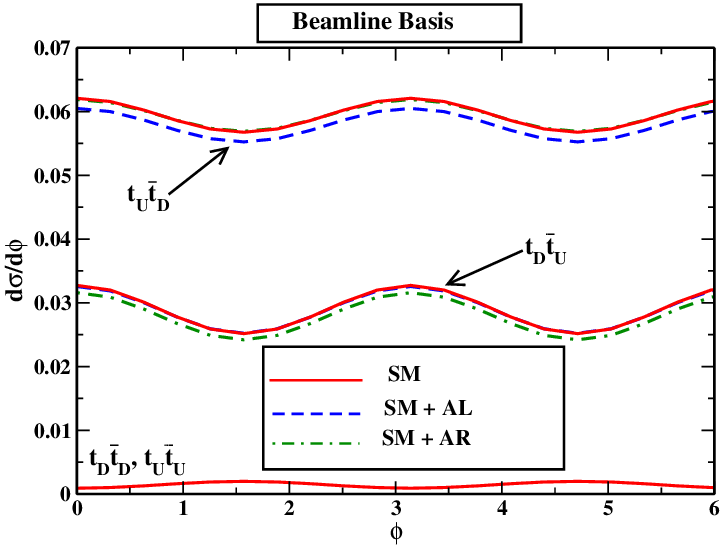}
\caption{}
\label{fig:beam_al}
\end{subfigure}\\[1ex]
\begin{subfigure}{\linewidth}
\centering
\vspace*{0.45cm}
\psfrag{SM}[b][b][0.75]{\bf{SM}}
\psfrag{SM + AL}[b][b][0.75]{\bf{SM +} $\bf{A^{'}_{L}}$}
\psfrag{SM + AR}[b][b][0.75]{\bf{SM +} $\bf{A^{'}_{R}}$}
\includegraphics[width=7.5cm, height=6cm]{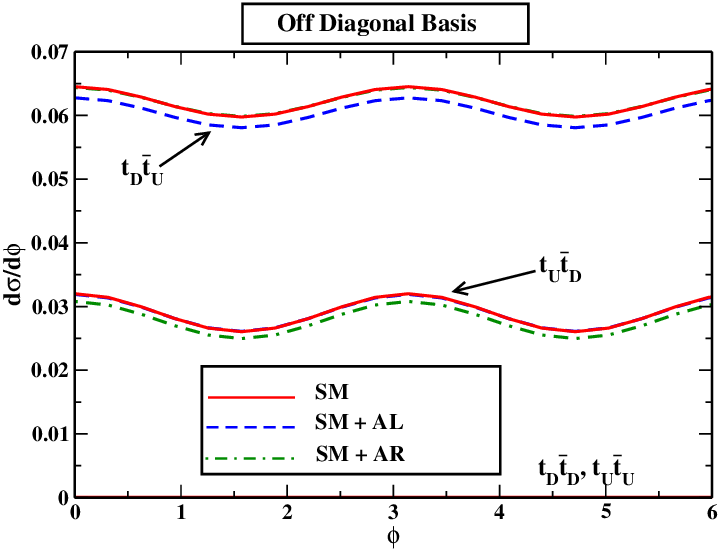}
\caption{}
\label{fig:offd_al}
\end{subfigure}
\caption{The azimuthal distribution of the top quark pairs in different final state polarization
at a 500 GeV linear collider for $P^T_{e^-}$ = 0.8, $P^T_{e^+}$ = 0.6 .  Different spin bases discussed
in the paper are considered  for SM and the case with either of the anomalous coupling $A^{'}_{L,R} = 10^{-7}$ GeV$^{-2}$
while keeping the value of others to be zero.}
\label{fig:azi_alar}
\end{figure}

The azimuthal distribution in the presence of $A^{'}_{L,R}$ along with the SM is shown in
Fig.~\ref{fig:azi_alar}. The analysis are performed by taking each time one of the anomalous couplings
to be 10$^{-7}$ GeV$^{-2}$, while the others are kept at zero. 
We have checked the linear approximation holds for this choice of
couplings.
We are here following the spirit of~\cite{hioki1}
for the chosen value of the anomalous coupling. As in this analysis the contribution from
$A^{'}_{L,R}$ and $B^{'}_{L,R}$ is considered at linear order only, therefore the effect due to new physics 
on the total cross section is from its interference with the SM couplings. This is in contrast with~\cite{hioki1},
where they consider new physics to quadratic order.
The deviation from SM in case of different bases
for different couplings varies depending upon the nature of the couplings. We note that
the couplings $B^{'}_{L,R}$, are always accompanied by $\sin\xi$, in case of the final state polarization
combination $t_U\bar{t}_U$ and $t_D\bar{t}_D$, Eq.~(\ref{eqn_vauu}). This reduces to zero 
in the HB, showing no effect in the azimuthal distribution.
                    
                    We note from Fig.~\ref{fig:azi_alar}, that the azimuthal distribution
for the  $t_U\bar{t}_U$ and $t_D\bar{t}_D$ final-state spin combination, is always the same in
case of all the bases. The behaviour in case of HB can be understood from the helicity amplitudes for the 
$t_U\bar{t}_U$ and $t_D\bar{t}_D$ final-state spin combination Appendix~\ref{hel_amp} in the presence of $V$ 
and $A$ interactions. They are the same apart from a minus sign. 
In case of the other bases BLB and SMOD, the amplitudes for these final-spin configurations 
are obtained by the action of the transformation matrix Eqs.~(\ref{equ:rot},~\ref{mat1}) resulting in the 
amplitudes which are also equal for $t_U\bar{t}_U$ and $t_D\bar{t}_D$
apart from a minus sign.  Therefore these particular spin
configurations always show the same behaviour in case of the three bases, for all the observables considered here.
                   
\subsection{Effect of Longitudinal Beam Polarization}

\vspace{0.3cm}        
\begin{figure}[htb]
\begin{subfigure}{.45\linewidth}
\centering
\psfrag{SM}[b][b][0.75]{\bf{SM}}
\psfrag{SM + AL}[b][b][0.75]{\bf{SM +} $\bf{A^{'}_{L}}$}
\psfrag{SM + AR}[b][b][0.75]{\bf{SM +} $\bf{A^{'}_{R}}$}
\psfrag{SM + BL}[b][b][0.75]{\bf{SM +} $\bf{B^{'}_{L}}$}
\psfrag{SM + BR}[b][b][0.75]{\bf{SM +} $\bf{B^{'}_{R}}$}
\psfrag{Pe = 0, Pp = 0}[b][b][0.75]{$\bf{P^L_{e^-}}$ \bf{= 0,} $\bf{P^L_{e^+}}$ \bf{= 0}}
\includegraphics[width=7.5cm, height=6cm]{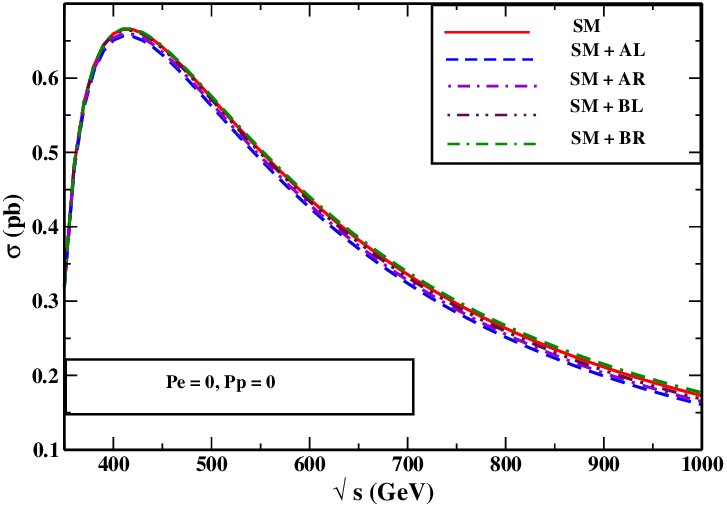}
\caption{}
\label{fig:sub1}
\end{subfigure}%
\begin{subfigure}{.45\linewidth}
\centering
\psfrag{SM}[b][b][0.75]{\bf{SM}}
\psfrag{SM + AL}[b][b][0.75]{\bf{SM +} $\bf{A^{'}_{L}}$}
\psfrag{SM + AR}[b][b][0.75]{\bf{SM +} $\bf{A^{'}_{R}}$}
\psfrag{SM + BL}[b][b][0.75]{\bf{SM +} $\bf{B^{'}_{L}}$}
\psfrag{SM + BR}[b][b][0.75]{\bf{SM +} $\bf{B^{'}_{R}}$}
\psfrag{Pe = -0.8, Pp = 0.6}[b][b][0.75]{$\bf{P^L_{e^-}}$ \bf{= -0.8,} $\bf{P^L_{e^+}}$ \bf{= 0.6}}
\includegraphics[width=7.5cm, height=6cm]{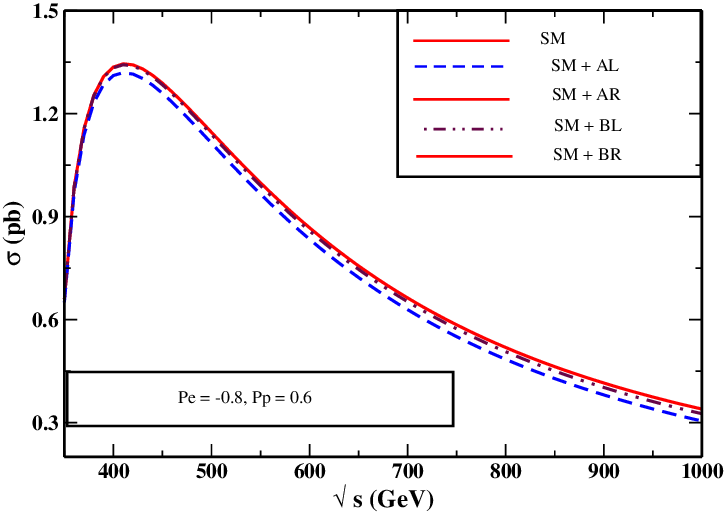}
\caption{}
\label{fig:sub2}
\end{subfigure}\\[1ex]
\begin{subfigure}{\linewidth}
\centering
\vspace*{0.75cm}
\psfrag{SM}[b][b][0.75]{\bf{SM}}
\psfrag{SM + AL}[b][b][0.75]{\bf{SM +} $\bf{A^{'}_{L}}$}
\psfrag{SM + AR}[b][b][0.75]{\bf{SM +} $\bf{A^{'}_{R}}$}
\psfrag{SM + BL}[b][b][0.75]{\bf{SM +} $\bf{B^{'}_{L}}$}
\psfrag{SM + BR}[b][b][0.75]{\bf{SM +} $\bf{B^{'}_{R}}$}
\psfrag{Pe = 0.8, Pp = -0.6}[b][b][0.75]{$\bf{P^L_{e^-}}$ \bf{= 0.8,} $\bf{P^L_{e^+}}$\bf{ = -0.6}}
\includegraphics[width=7.5cm, height=6cm]{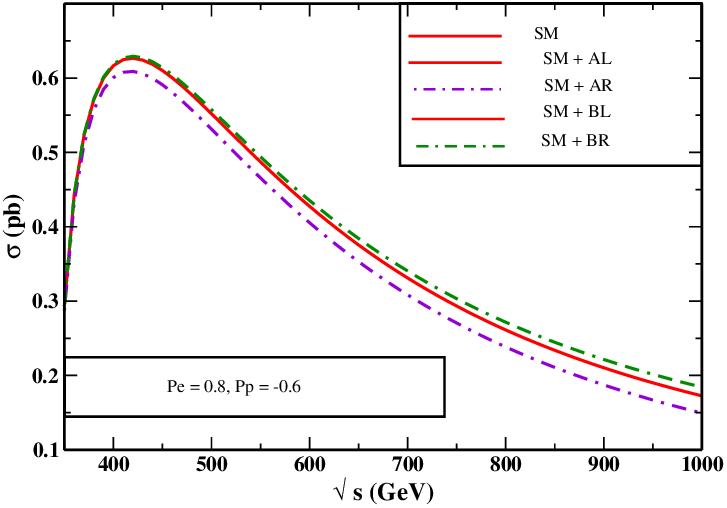}
\caption{}
\label{fig:sub3}
\end{subfigure}
\caption{The total cross section in case of different beam polarizations for $t\bar{t}$ pair
production as a function of center of mass energy $\sqrt{s}$ for SM and in the presence of different anomalous couplings
each with a value of $10^{-7}$ GeV$^{-2}$ contributing individually, keeping the values of other couplings to be zero.}
\label{fig:cross}
\end{figure}            

 The new physics in the form of non-standard interactions of $V$ and $A$
type, can also be studied with unpolarized and
longitudinally polarized beams. For a complete analysis, we carry out a detailed study
of the behaviour due to these non-standard interactions ($V$, $A$) in the presence of both unpolarized and longitudinally polarized beams. 
In case of LP, the differential 
cross section is obtained in a straight forward manner:
\begin{eqnarray}
\frac{d\sigma(e^+e^-\rightarrow t_I\bar{t}_J)}{d\cos \theta}=\frac{\pi}{2}\left((1+P^L_{e^-})(1-P^L_{e^+})|T_{RLIJ}|^2+
(1-P^L_{e^-})(1+P^L_{e^+})|T_{LRIJ}|^2\right),
\end{eqnarray}

where $P^L_{e^-}$ and $P^L_{e^+}$ are the LP of the electron and positron beam respectively. The amplitudes
$T_{RLIJ}$ and $T_{LRIJ}$ are defined in Appendix.~\ref{hel_amp}.
We show in Fig.~\ref{fig:cross} the total unpolarized cross section along with the polarized one for different cases
of beam polarization, as a function of $\sqrt{s}$ in case of $t\bar{t}$ pair production, for SM and in the presence of new physics. 
We have taken for the NP, the couplings with a value of $10^{-7}$ GeV$^{-2}$ as in the case of TP.
 For the case of unpolarized beams Fig.~\ref{fig:sub1}, the deviation due to the anomalous couplings
$B^{'}_{L,R}$ from the SM is not much pronounced compared to $A^{'}_{L,R}$. At $\sqrt{s}$ = 500 GeV, the deviation due to $B^{'}_{L,R}$
is about 0.4\% from SM, whereas $A^{'}_{L,R}$ produces 2\%  deviation.  The implementation of beam polarization,
with $P^L_{e^-}$ = -0.8, $P^L_{e^+}$ = 0.6, as shown in
Fig.~\ref{fig:sub2}, increases the sensitivity to $A^{'}_{L}$ and $B^{'}_{L}$, 
along with
a increase in statistics. Fig.~\ref{fig:sub3} shows the polarization 
combination $P^L_{e^-}$ = 0.8, $P^L_{e^+}$ = -0.6,
with an enhanced sensitivity to the anomalous couplings $A^{'}_{R}$ and $B^{'}_{R}$.
            
 The angular correlation between the $t\bar{t}$ in the case of the SM, is the
best for the SMOD.  We next describe an asymmetry where this angular  correlation between the 
final state products can be observed. This was earlier considered in the case of hadron colliders~\cite{Head},~\cite{hep-ph/9512292}
based upon the asymmetry in the number of like spin to unlike spin $t\bar{t}$ pairs produced.

\begin{eqnarray}\label{asym3}
A_{t\bar t}=\frac{(N_{UU}+N_{DD})-(N_{DU}+N_{UD})}{N_{UU}+N_{DD}+N_{DU}+N_{UD}}
\end{eqnarray}

 $N_{ij}$ ($i,j = U,D$) in Eq.~(\ref{asym3}) denotes the number of
events for the top and anti-top spin combinations $t_i$ and $\bar{t}_j$.
Table~\ref{table_asyml} shows correlations for different spin bases in
the different cases
of LP for SM along with new physics couplings of vector and axial-vector type. The correlation
measured by the asymmetry $A_{t \bar t}$ is sensitive to the choice of the bases, along with the
initial beam polarization. 
At a 500 GeV unpolarized linear collider, about 83\% of the final state pairs have opposite helicity,
whereas 17\% have the same helicity. The correlation in Eq.~(\ref{asym3}) 
is seen to be
about 66\% from the results in Table~\ref{table_asyml}
for the helicity basis.  
Although SMOD is the best basis for observing the correlation,
it is not sensitive in distinguishing contributions from new physics. It can be seen from the Table,
that for different cases of LP the HB and BLB are more sensitive to new physics couplings compared to SMOD.

\begin{table}[htb]
\begin{center}
\begin{tabular}{|c|c|c|c|c|}
\hline
Couplings &Spin Bases &$P^L_{e^-}$ = 0.8, $P^L_{e^+}$ = -0.6 &$P^L_{e^-}$ = -0.8, $P^L_{e^+}$ = 0.6 &$P^L_{e^-}$ = 0, $P^L_{e^+}$ = 0 \\ \hline
 & HB &-0.6707 &-0.6516 &-0.6578 \\
SM &BLB &-0.9608 &-0.9247 &-0.9364 \\
 &SMOD &-0.9940 &-0.9999 &-0.9980 \\ \hline
 &HB &-0.6707 &-0.6525 &-0.6586 \\
SM + $A^{'}_L$ &BLB &-0.9610 &-0.9270 &-0.9383 \\
 &SMOD &-0.9940 & -0.9999 &-0.9980 \\ \hline
  &HB &-0.6727 &-0.6516 &-0.6583 \\  
SM + $A^{'}_R$ &BLB &-0.9638 & 0.9247 &-0.9370 \\
 &SMOD &-0.9928 &-0.9999 &-0.9977 \\ \hline
 &HB &-0.6706 &-0.6496 &-0.6564 \\
SM + $B^{'}_L$ &BLB &-0.9605 &-0.9198 &-0.9331 \\
 &SMOD &-0.9940 &-0.9999 &-0.9980 \\ \hline
 &HB  &-0.6739 &-0.6516 &-0.6589 \\
SM + $B^{'}_R$ &BLB &-0.9655 &-0.9247 & -0.9380 \\
&SMOD &-0.9922 &-0.9999 &-0.9974 \\ \hline    
\end{tabular}
\caption{The asymmetry $A_{t \bar t}$, Eq.~(\ref{asym3}) measuring the strength of the correlation in different spin bases for
different beam polarizations in case of SM and other non-standard interactions of $V$ and $A$ type at a centre of mass energy of 500 GeV.}
\label{table_asyml}
\end{center}
\end{table}

For purpose of completeness we obtain 90\% CL limits on the $V$ and $A$ couplings for realistic beam polarizations and typical
integrated luminosity. The limits obtained for the couplings in case of different bases for different beam
polarization is shown in Table~\ref{tab:limit}. It can be seen from the Table that the limits obtained in this case are not
very competitive. These interactions being similar to SM, it is very difficult to isolate their signatures unlike $S$ and $T$.  
Therefore it is very difficult to see the  new physics signatures of $V$ and $A$ unless we go to higher centre of mass energy and higher luminosity.

\begin{table}[htb]
\begin{center}
\begin{tabular}{|c|c|c|c|c|}
\hline
Couplings &Spin Bases &$P^L_{e^-}$ = 0.8, $P^L_{e^+}$ = -0.6 &$P^L_{e^-}$ = -0.8, $P^L_{e^+}$ = 0.6 &$P^L_{e^-}$ = 0, $P^L_{e^+}$ = 0 \\ \hline
  &HB & 0.0051 GeV$^{-2}$ & 0.0035 GeV$^{-2}$& 0.0050 GeV$^{-2}$ \\
 $A^{'}_L$ &BLB & 0.0037 GeV$^{-2}$& 0.0026 GeV$^{-2}$& 0.0036 GeV$^{-2}$\\
 &SMOD & 0.0031 GeV$^{-2}$& 0.0022 GeV$^{-2}$&0.0031 GeV$^{-2}$\\ \hline
  &HB & 0.0040 GeV$^{-2}$& 0.0034 GeV$^{-2}$& 0.0050 GeV$^{-2}$  \\  
 $A^{'}_R$ &BLB & 0.0033 GeV$^{-2}$& 0.0356 GeV$^{-2}$ & 0.0035 GeV$^{-2}$  \\
 &SMOD & 0.0031 GeV$^{-2}$& 0.0014 GeV$^{-2}$& 0.0030 GeV$^{-2}$  \\ \hline
 &HB & 0.0031 GeV$^{-2}$& 0.0022 GeV$^{-2}$& 0.0031 GeV$^{-2}$  \\
 $B^{'}_L$ &BLB & 0.0018 GeV$^{-2}$& 0.0012 GeV$^{-2}$& 0.0017  GeV$^{-2}$  \\
 &SMOD & 0.0031 GeV$^{-2}$& 0.0022 GeV$^{-2}$ & 0.0031 GeV$^{-2}$ \\ \hline
 &HB & 0.0031 GeV$^{-2}$& 0.0022 GeV$^{-2}$& 0.0031 GeV$^{-2}$  \\
 $B^{'}_R$ &BLB & 0.0026 GeV$^{-2}$& 0.0012 GeV$^{-2}$& 0.0021 GeV$^{-2}$  \\
&SMOD & 0.0035 GeV$^{-2}$& 0.0067 GeV$^{-2}$& 0.0038  GeV$^{-2}$ \\ \hline    
\end{tabular}
\caption{90\% CL limit obtained on various coupling from the asymmetry $A_{t \bar t}$ Eq.~(\ref{asym3}), at
$\sqrt{s}$ = 500 GeV with an integrated luminosity of 500 fb$^{-1}$ for different cases of LP.}
\label{tab:limit}
\end{center}
\end{table}

\vspace{0.2cm}
\begin{figure}[htb]
\begin{subfigure}{.45\linewidth}
\centering
\psfrag{SM}[b][b][0.75]{\bf{SM}}
\psfrag{SM + AL}[b][b][0.75]{\bf{SM +} $\bf{A^{'}_{L}}$}
\psfrag{SM + BL}[b][b][0.75]{\bf{SM +} $\bf{B^{'}_{L}}$}
\includegraphics[width=7.5cm, height=6cm]{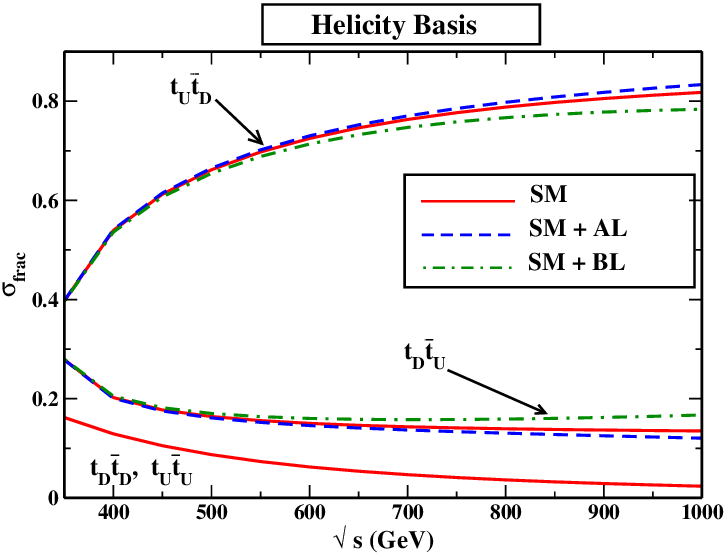}
\caption{}
\label{fig:lr_frac1}
\end{subfigure}%
\begin{subfigure}{.45\linewidth}
\centering
\psfrag{SM}[b][b][0.75]{\bf{SM}}
\psfrag{SM + AL}[b][b][0.75]{\bf{SM +} $\bf{A^{'}_{L}}$}
\psfrag{SM + BL}[b][b][0.75]{\bf{SM +} $\bf{B^{'}_{L}}$}
\includegraphics[width=7.5cm, height=6cm]{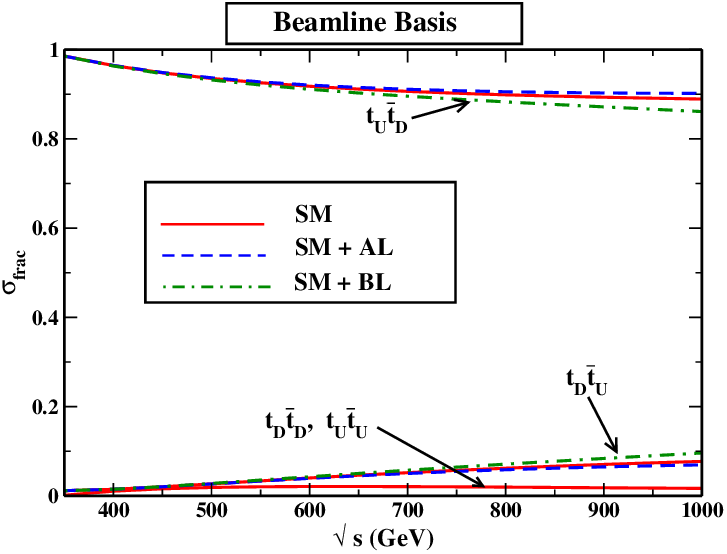}
\caption{}
\label{fig:lr_frac2}
\end{subfigure}\\[1ex]
\begin{subfigure}{\linewidth}
\centering
\vspace*{0.45cm}
\psfrag{SM}[b][b][0.75]{\bf{SM}}
\psfrag{SM + AL}[b][b][0.75]{\bf{SM +} $\bf{A^{'}_{L}}$}
\psfrag{SM + BL}[b][b][0.75]{\bf{SM +} $\bf{B^{'}_{L}}$}
\includegraphics[width=7.5cm, height=6cm]{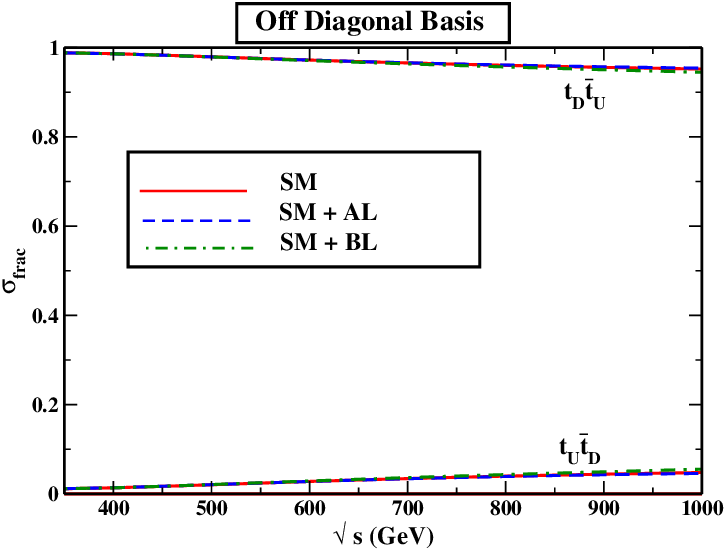}
\caption{}
\label{fig:lr_frac3}
\end{subfigure}
\caption{The fraction of the $t\bar{t}$ pair
production as a function of center of mass energy $\sqrt{s}$ for SM and in the presence of $A^{'}_{L}$, $B^{'}_{L}$
each with a value of $10^{-7}$ GeV$^{-2}$ contributing individually, keeping the values of other couplings to be zero. The 
three bases discussed here are considered, with an initial beam polarization of $P^L_{e^-}$ = -0.8, $P^L_{e^+}$ = 0.6.}
\label{fig:frac_lr}
\end{figure}       

\begin{figure}[htb]
\begin{subfigure}{.45\linewidth}
\centering
\psfrag{SM}[b][b][0.75]{\bf{SM}}
\psfrag{SM + AR}[b][b][0.75]{\bf{SM +} $\bf{A^{'}_{R}}$}
\psfrag{SM + BR}[b][b][0.75]{\bf{SM +} $\bf{B^{'}_{R}}$}
\includegraphics[width=7.5cm, height=6cm]{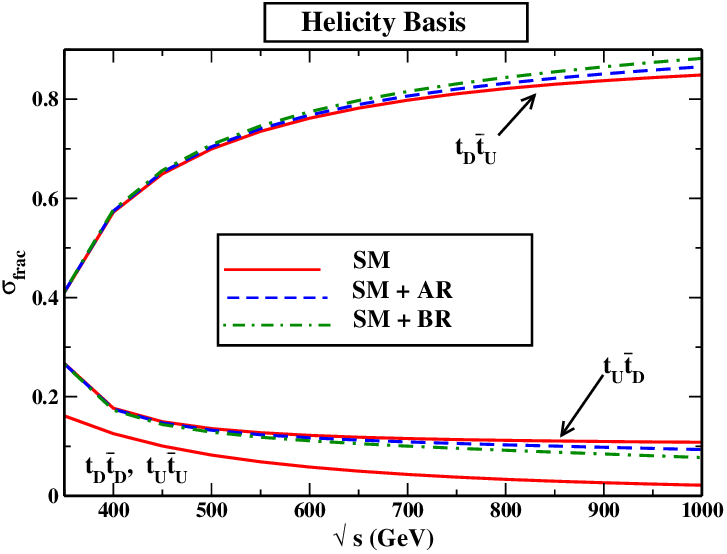}
\caption{}
\label{fig:rl_frac1}
\end{subfigure}%
\begin{subfigure}{.45\linewidth}
\centering
\psfrag{SM}[b][b][0.75]{\bf{SM}}
\psfrag{SM + AR}[b][b][0.75]{\bf{SM} + $\bf{A^{'}_{R}}$}
\psfrag{SM + BR}[b][b][0.75]{\bf{SM} + $\bf{B^{'}_{R}}$}
\includegraphics[width=7.5cm, height=6cm]{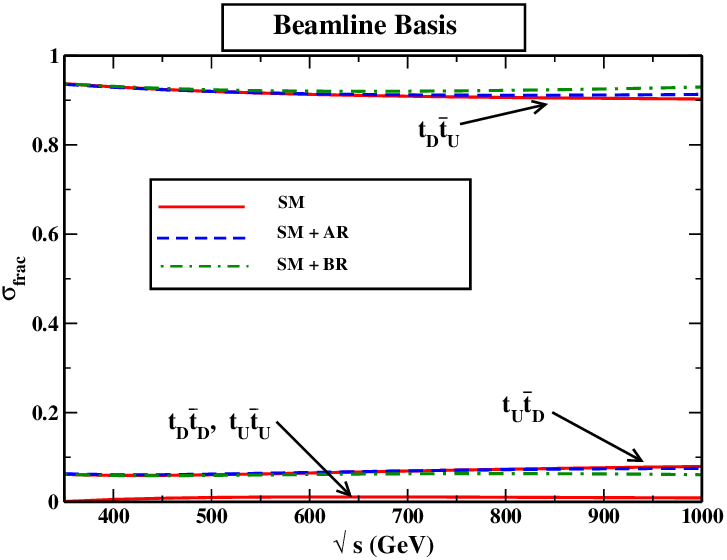}
\caption{}
\label{fig:rl_frac2}
\end{subfigure}\\[1ex]
\begin{subfigure}{\linewidth}
\centering
\vspace*{0.45cm}
\psfrag{SM}[b][b][0.75]{\bf{SM}}
\psfrag{SM + AR}[b][b][0.75]{\bf{SM} + $\bf{A^{'}_{R}}$}
\psfrag{SM + BR}[b][b][0.75]{\bf{SM} + $\bf{B^{'}_{R}}$}
\includegraphics[width=7.5cm, height=6cm]{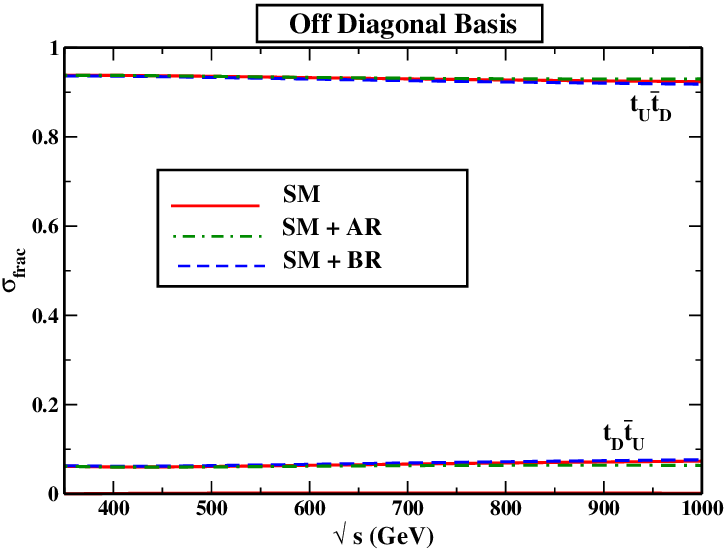}
\caption{}
\label{fig:rl_frac3}
\end{subfigure}
\caption{The fraction of the $t\bar{t}$ pair
production as a function of center of mass energy $\sqrt{s}$ for SM and in the presence of $A^{'}_{R}$, $B^{'}_{R}$
each with a value of $10^{-7}$ GeV$^{-2}$ contributing individually, keeping the values of other couplings to be zero. The 
three bases discussed here are considered, with an initial beam polarization of $P^L_{e^-}$ = 0.8, $P^L_{e^+}$ = -0.6.}
\label{fig:frac_rl}
\end{figure}

  We have also considered the fraction of $t\bar{t}$ pairs produced in
different polarization states versus the $e^+e^-$
c.m. energy in case of different spin bases. The presence of the anomalous couplings
along with SM is considered, for different cases of initial LP.  The fraction 
of $t\bar t$ pairs produced in a spin combination $s_ts_{\bar t}$ 
is defined as 
\begin{equation}
\sigma_{\rm{frac}}=\frac{\sigma(e^+e^- \rightarrow t_{s_t}\bar{t}_{s_{\bar{t}}})}{\sigma_{\rm{tot}}},
\end{equation}
where $\sigma_{\rm{tot}}$ is the total cross section for unpolarized
$t$,$\bar t$,
with possible inclusion of anomalous couplings.
The results are shown in Figs.~\ref{fig:frac_lr},~\ref{fig:frac_rl} for realistic initial beam polarizations of 
$P^L_{e^-}$ = -0.8, $P^L_{e^+}$ = 0.6 
and $P^L_{e^-}$ = 0.8, $P^L_{e^+}$ = -0.6.

  In the SMOD shown in Figs.~\ref{fig:lr_frac3},~\ref{fig:rl_frac3} the polarization of $t\bar{t}$ states
compared to the BLB Figs.~\ref{fig:lr_frac2},~\ref{fig:rl_frac2} and HB Figs.~\ref{fig:lr_frac3},~\ref{fig:rl_frac3}
is more. With the  left handed electrons initial beam polarization, the spin configuration $t_D\bar{t}_U$ gives the dominant
contribution in the SMOD, whereas for right handed initial beam polarization the dominant configuration comes
from $t_U\bar{t}_D$. This behaviour is similar to that observed in case
of the asymmetry $A_{t\bar t}$ measuring the
spin correlation in different bases. The effect of the new physics is most dominantly seen in the HB and the least in the
SMOD.  These additional effects are mainly from the interference term between the NP and SM.
In the SMOD, since one of the final-state spin configuration is dominant it forces the new physics contributing
to the interference term to be in the same spin configuration. The helicity basis on the other hand  treats the final
state spin configurations with opposite helicities almost equally allowing the new physics
to freely interfere with the SM couplings, leading to deviations compared to the SM predictions.  Thus in
case of LP or unpolarized beams both in the case of work~\cite{Lin} and in the present work the HB and BLB
works better compared to SMOD, in search for new physics.

\vspace{0.2 cm}

\section{Discussions and Conclusions}\label{conclusions}

    The top quark due to its large mass compared to other quarks has been an interesting
tool to look for significant deviations from the SM. Because of its short lifetime the top quark decays
before hadronization, with its spin surviving and showing its effect through the distribution of the
decay products. The measurement of the spin can be done through the analysis of the decay products. At the ILC
the $t\bar{t}$ pairs will be produced in large numbers, and will be an ideal tool to look for BSM
effects in top quarks couplings. The degree
of polarization can be changed by tuning the initial beams polarization.  The spin correlation in top quark production
is therefore an interesting issue in top quark physics.

     PS~\cite{Parke} had suggested different spin bases to study spin correlations. We have presented 
a new and simpler derivation, which accounts the most general physics of
$S$, $P$, $V$, $A$ and $T$ types. The reason for these new studies is the
realization that the number of like spin and unlike spin top quarks can be made significantly different by an appropriate
choice of spin bases. Experimentally the $t\bar{t}$ spin correlation is measured by analyzing the angular distribution
of the t and $\bar{t}$ decay products. As the top quark decays through left handed weak force, it analyses its own polarization 
through its decay products. To make a practical use of the bases defined here, the direction of the charged lepton momentum
in a top leptonic decay must be observed, giving an indication of the top polarization direction. The angular distribution
of the decay products in top quark decays, according to the polarization of the parent top quark is given by:
\begin{equation}
\frac{1}{\Gamma}\frac{d\Gamma}{d\cos\theta_i}=\frac{1}{2}(1\pm A_i \cos\theta_i)
\end{equation}
where the $\pm$ sign in front of $A_i$ is used for right-handed and left-handed quarks respectively. The angle $\theta_i$
is the angle between the spin quantization axis and the momentum of the decay particle in top quark rest frame. $A_i$ is
defined as the spin analysing power coefficient equal to 1 for the charged lepton or the down type quark.
The SMOD
discussed in~\cite{Parke}, has the feature that the production cross section of like spin states is almost negligible. With
the initial polarization of left (right) handed electron beam, the spin configuration $t_U\bar{t}_D$ ($t_D\bar{t}_U$)
gives the dominant cross section. Therefore observation of a sizable event rate in the like spin states
or a significant deviation from the dominant cross section in this basis will account for new physics signals.

    In this work we have looked for new physics signatures of the scalar ($S$) and tensor ($T$) type along with the
vector ($V$) and axial-vector ($A$) type of interactions. The signatures of these $S$ and $T$ 
interactions can only be probed at  linear order with TP of one 
or both the beams. We check the sensitivity of the bases discussed earlier, to $S$ and $T$ interactions, with one of the beams
having TP. This eliminates the $\phi$ contribution from SM and other interactions of $V$ and $A$ type. Each
of the bases bears a different signature. The BLB is found to be the most preferred basis in the
presence of TP, as it receives contribution from all the final state spin configuration. 
We then consider some asymmetries, where the SM contribution vanish.
Therefore any sizable observation will confirm the signatures of $S$ and $T$ type of physics.
Here too BLB is most sensitive to these NP effects, followed by SMOD and HB. Thus as discussed
in Sec.~\ref{distribution} CP violation as a probe of new physics through non vanishing Im $S$ and
Im $T$, is most likely to show up in the BLB analysis. We have also used an asymmetry analogous to the
one considered in~\cite{BAMPSDR} to obtain 90\% CL on the couplings with realistic polarization and luminosity.

    We have also looked for new physics in the form of $V$ and $A$ type of interactions, with all possible types 
of initial-beam polarizations (TP and LP). In case of TP, both the initial beams have to be polarized. The analysis
in case of TP shows similar results as in $S$ and $T$ scenario, with the BLB being the most sensitive to NP, and
the HB receiving the smallest amount of contributions from NP.  For unpolarized and longitudinal
polarized beams we have studied the correlation asymmetry for the different bases and have quoted the 
results in Table~\ref{table_asyml}. The asymmetry for the purpose of 90\% CL is
not very competitive. In case of LP, the result is contrary to TP, with the HB receiving the
significant amount of contribution from NP, and SMOD being the least sensitive. In the ILC, with the
planned polarization programme, BLB is the best in the presence of TP, followed by SMOD and HB, for 
the NP considered here. Similarly HB is the best in presence of
LP followed by BLB and SMOD for the study of NP. 

While in the past we have taken the approach of eliminating the sensitivity with realistic degrees of polarization
and integrated luminosity $\mathcal{L}$, here we take a complementary approach of assuming some \textquoteleft realistic'
values for BSM couplings to study the size of the signal. This enables us to clearly establish that the sensitivity levels established 
in~\cite{BAMPSDR} can be significantly improved. Since a realistic study will necessarily involve detector simulation studies,
our approach provides a clear analytical picture of the scale to which BSM physics in the sector considered here can be probed.

\vspace{0.5cm}
\noindent Acknowledgments: SDR gratefully acknowledges hospitality at
the Centre for High Energy Physics, Indian Institute of Science,
Bangalore, where this work was initiated. He also acknowledges financial support
from the Department of Science and Technology, India, in the form of a
J.C. Bose National Fellowship, grant no. SR/SB/JCB-42/2009.

\appendix
\section{Helicity Amplitudes}\label{hel_amp}

The helicity amplitudes for the process $e^+e^- \rightarrow t\bar{t}$ are defined below. They are the same as considered in
Ref.\cite{Grzad}, with the normalization factor taken care of.  The amplitudes
of $e^-$, $e^+$, $t$ and $\bar{t}$ are defined in the order $T_{LRIJ}$, where $L$ denotes the left-handed electron beam $e^-_L$, 
$R$ for right-handed positron beam $e^+_R$, and $IJ$ denotes the different final-state combinations of $t\bar{t}$,
i.e. $DD$, $DU$, $UD$ and $UU$.  Similarly $T_{RLIJ}$ denotes the right-handed electron beam $e^-_R$ and left-handed
positron beam $e^+_L$.

For the helicity-conserving interactions, the amplitudes are as follows:

\begin{eqnarray}\label{va_hel}
&&T_{LRUU} =B_1 A_L m_t \sin \theta  \\ \nonumber
&&T_{LRUD} =B_1 (E A_L + k B_L) (1 + \cos \theta)  \\ \nonumber
&&T_{LRDU} =-B_1 (E A_L - k B_L) (1 - \cos \theta)  \\ \nonumber
&&T_{LRDD} =-B_1 A_L m_t \sin \theta \\ \nonumber
&&T_{RLUU} =B_1 A_R m_t \sin \theta   \\ \nonumber
&&T_{RLUD} =-B_1 (E A_R + k B_R) (1 - \cos \theta)  \\ \nonumber
&&T_{RLDU} =B_1 (E A_R - k B_R) (1 + \cos \theta)   \\ \nonumber
&&T_{RLDD} =-B_1 A_R m_t \sin \theta.   \\  \nonumber
\end{eqnarray}

All the expressions above have the normalization factor $B_1$ defined as  $i\sqrt{3 \beta \alpha^2/4}$.
$E$ is the  beam  energy $\sqrt{s}/2$ and $k = E\beta$, where $\beta =\sqrt{1-4m_t^2/s}$. The amplitudes
in the presence of vector and axial vector type four-Fermi operator effects, has the same form as those above. 
In our analyses, we have considered the effect of new physics only through its interference with the SM amplitudes.

Similarly, for the helicity-violating interactions, the amplitudes are:

\begin{eqnarray}\label{st_hel}
&&T_{LLUU} =A_1( (E+k) (S_{LL} - 2 T_{LL} \cos \theta) - S_{LR} (E-k)) \\ \nonumber
&&T_{LLUD} = 2 A_1  T_{LL} m_t \sin \theta \\ \nonumber
&&T_{LLDU} =2 A_1  T_{LL} m_t \sin \theta \\ \nonumber
&&T_{LLDD} =A_1( (E-k) (S_{LL} + 2 T_{LL} \cos \theta) - S_{LR} (E+k)) \\ \nonumber
&&T_{RRUU} =A_1( (E-k) (S_{RR} + 2 T_{RR} \cos \theta) - S_{RL} (E-k)) \\ \nonumber
&&T_{RRUD} =-2 A_1 T_{RR} m_t \sin \theta \\ \nonumber
&&T_{RRDU} =-2 A_1 T_{RR} m_t \sin \theta \\  \nonumber
&&T_{RRDD} =A_1( (E+k) (S_{RR} - 2 T_{RR} \cos \theta) - S_{RL} (E-k)). \\ \nonumber
\end{eqnarray} 

The normalization factor for the above amplitudes $A_1$ is defined as $i \sqrt{3 \beta/64 \pi^2}$.
The expressions for the scalar $S$ and tensor $T$ operators are as follows:

\begin{eqnarray}
S_{RR} =  \rm{Re} S + i \rm{Im} S,  \ \ \    S_{LL} = \rm{Re} S - i \rm{Im} S, \ \ \ 
T_{RR} = \rm{Re} T + i \rm{Im} T,   \ \ \     T_{LL} = \rm{Re} T - i \rm{Im} T
\end{eqnarray}

\end{document}